\documentclass[aps,prc,10pt,twocolumn,amsmath,amssymb,superscriptaddress]{revtex4-2}
\usepackage{graphicx}
\usepackage{color}
\usepackage{lineno}
\newcommand{\be}{\begin{equation}}
\newcommand{\ee}{\end{equation}}
\newcommand{\beq}{\begin{eqnarray}}
\newcommand{\eeq}{\end{eqnarray}}
\newcommand{\ba}{\begin{array}}
\newcommand{\ea}{\end{array}}

\begin{document}

\title{Single-pion contribution to the Gerasimov--Drell--Hearn sum rule and related integrals}

\vspace{3mm}
\date{\today}

\author{\mbox{Igor~Strakovsky}}
\altaffiliation{Corresponding author: \texttt{igor@gwu.edu}}
\affiliation{Institute for Nuclear Studies, Department of Physics, 
    The George Washington University, Washington, DC 20052, USA}

\author{\mbox{Simon~\v{S}irca}}
\affiliation{Faculty of Mathematics and Physics, University of Ljubljana, 
    1000 Ljubljana, Slovenia}

\author{\mbox{William~J.~Briscoe}}
\affiliation{Institute for Nuclear Studies, Department of Physics, The
        George Washington University, Washington, DC 20052, USA}

\author{\mbox{Alexandre~Deur}}
\affiliation{Thomas Jefferson National Accelerator Facility, Newport News, Virginia 23606, USA}

\author{\mbox{Axel~Schmidt}}
\affiliation{Institute for Nuclear Studies, Department of Physics, 
    The George Washington University, Washington, DC 20052, USA}

\author{\mbox{Ron~L.~Workman}}
\affiliation{Institute for Nuclear Studies, Department of Physics, 
    The George Washington University, Washington, DC 20052, USA}

\noaffiliation

\begin{abstract}
Phenomenological amplitudes obtained in partial-wave analyses (PWA) of single-pion photoproduction are used to evaluate the contribution of this process to the Gerasimov--Drell--Hearn (GDH), Baldin and Gell-Mann--Goldberger--Thirring (GGT) sum rules, by integrating up to $2~\mathrm{GeV}$ in photon energy.  Our study confirms that the single-pion contribution to all these sum rules converges even before the highest considered photon energy, but the levels of saturation are very different in the three cases. Single-pion production almost saturates the GDH sum rule for the proton, while a large fraction is missing in the neutron case. The Baldin integrals for the proton and the neutron are both saturated to about four fifths of the predicted total strength.  For the GGT sum rule, the wide variability in predictions precludes any definitive statement. 
\end{abstract}

\maketitle

\clearpage
\section{Introduction}

The study of polarized lepton scattering off polarized nucleons provides information on the spin composition of the nucleon.  In the real-photon limit of the lepton-nucleon interaction, the process can be characterized
either in terms of static nucleon quantities or in terms of integrals of cross-sections with various weights, resulting in specific ``sum rules''.  The relevant quantities of interest are $\sigma_{3/2}$ and $\sigma_{1/2}$, 
the photon-nucleon total absorption cross-sections for circularly polarized photons on longitudinally polarized nucleons, with total helicity $3/2$ and $1/2$, respectively, from which either a difference, $\Delta\sigma = \sigma_{3/2} - \sigma_{1/2}$, or sum, $\sigma_\mathrm{tot} = \sigma_{3/2} + \sigma_{1/2}$, can be constructed. In this paper, we discuss three sum rules in which $\Delta\sigma$ and $\sigma_\mathrm{tot}$ are the crucial quantities entering the integrals with different weights. Comprehensive reviews of the status of the Gerasimov--Drell--Hearn (GDH) sum rule, the first to be discussed in this paper, and related integrals are given in Refs.~\cite{Drechsel:2004ki,Schumacher:2005an,Helbing:2006zp}.

\section{Sum rules involving $\Delta\sigma$ and $\sigma_\mathrm{tot}$}

The GDH sum rule, formulated in the 1960s~\cite{Gerasimov:1965et,Drell:1966jv}, rests upon fundamental physics principles (Lorentz invariance, gauge invariance, crossing symmetry, rotational invariance, causality, and unitarity) and unsubtracted dispersion relations applied to the forward Compton amplitude (see also Ref.~\cite{Lapidus:1962} for an earlier related work). 
Because of its fundamental character, the GDH sum rule requires experimental verification which has been awaiting technical developments that have only recently been attained.  

The GDH integral relates the proton (neutron) anomalous magnetic moments $\kappa_p = (\mu_p - \mu_N)/\mu_N \approx 1.793$ where $\mu_N$ is the nuclear magneton ($\kappa_n = \mu_n/\mu_N \approx -1.913$) to the integral of $\Delta\sigma$ weighted by the photon energy 
in the laboratory frame, $E_\gamma$: 
\begin{equation}
	I_\mathrm{GDH} = \int_{E_\gamma^\mathrm{thr}}^{\infty}\frac{\Delta\sigma}{E_\gamma}~dE_\gamma = \frac{2\pi^2\alpha}{M^2} \kappa^2 \>.
    \label{eq:GDH}
\end{equation}
Here $E_\gamma^\mathrm{thr}$ is the photon energy corresponding to the pion photoproduction threshold, $\alpha = e_0^2/4\pi$ is the fine-structure constant, and $M$ is the nucleon mass. 

For the proton and neutron, all static quantities appearing in Eq.~(\ref{eq:GDH}) are known very precisely.  For instance, the most recent value of the proton magnetic moment $\mu_p = 2.792847350(7)(6)~\mu_N$ was obtained by the BASE Collaboration and outperforms previous Penning trap measurements in terms of precision by a factor of about 760~\cite{BASE:2014drs}.  A similarly precise value for the neutron, $\mu_n = -1.91304272(45)~\mu_N$, was obtained by the ISOLDE Collaboration~\cite{ISOLDE:1999dkt} by measuring the magnetic moment of $^{11}\mathrm{Be}$ and detecting nuclear magnetic resonance signals in a beryllium crystal lattice. Consequently, the uncertainties on the nominal value of the GDH integral are also very small: one expects $204.784482(35)~\mu$b and $232.25159(13)~\mu$b for the proton and the neutron, respectively, by using the most recent evaluations of the fine-structure constant, measurements of the proton and neutron masses, and their magnetic moments~\cite{ParticleDataGroup:2020ssz}, with their uncertainties added quadratically. In contrast, constraints of $I_\mathrm{GDH}$ from photoproduction data are far less stringent.

The Baldin sum rule~\cite{Baldin:1960,Lapidus:1963} relates the sum of the electric, $\alpha$, and magnetic, $\beta$, polarizabilities of the nucleon to the total photoabsorption cross section $\sigma_\mathrm{tot}$,
\begin{eqnarray}
	I_\mathrm{Baldin} 
    = \frac{1}{4 \pi^2} \int_{E_\gamma^\mathrm{thr}}^{\infty} 
	\frac{\sigma_\mathrm{tot}}{E_\gamma^2}~dE_\gamma
    = \alpha + \beta \>.
    \label{eq:Baldin}
\end{eqnarray}
As pion photoproduction does not allow for a model-independent separation of electric and magnetic polarizabilities, we shall compare our results for the sum of $\alpha + \beta$, determined from Compton scattering. The PDG average for $\alpha + \beta$ is ($14.2\pm 0.5)\times 10^{-4}\,\mathrm{fm}^3$ for the proton and $(15.5\pm 1.6)\times 10^{-4}\,\mathrm{fm}^3$ for the neutron. 

The third sum rule, introduced by Gell-Mann--Goldberger--Thirring (GGT)~\cite{Gell-Mann:1954wra,Gell-Mann:1954ttj}, again involves $\Delta\sigma$ in the integrand, and results in the forward spin polarizability $\gamma_0$:
\begin{equation}
	I_\mathrm{GGT} = -\frac{1}{4 \pi^2} \int_{E_\gamma^\mathrm{thr}}^{\infty}\frac{\Delta\sigma}{E_\gamma^3}~dE_\gamma~ = \gamma_0 \>.
    \label{eq:ForPol}
\end{equation}

The sum rules considered in this paper call for both $\Delta\sigma$ and $\sigma_\mathrm{tot}$, and it is important to note that
at energies far above the nucleon resonance region (not far beyond invariant masses of several $\mathrm{GeV}$ (see Fig.~\ref{fig:sigtot} (bottom)) the unpolarized total photoabsorption cross-section, $\sigma_\mathrm{tot}$, appears to {\sl rise indefinitely\/}; in Regge theory, this rise can be explained in terms of processes involving pomeron exchange (see below), but it results in a non-convergent integral of $\sigma_\mathrm{tot}$, which  puts into question the asymptotic behavior of $\Delta\sigma$ as well.


\section{Previous evaluations}

\begin{figure}[hbtp]
\vspace{0.4cm}
\centering
{
    \includegraphics[width=0.43\textwidth,keepaspectratio]{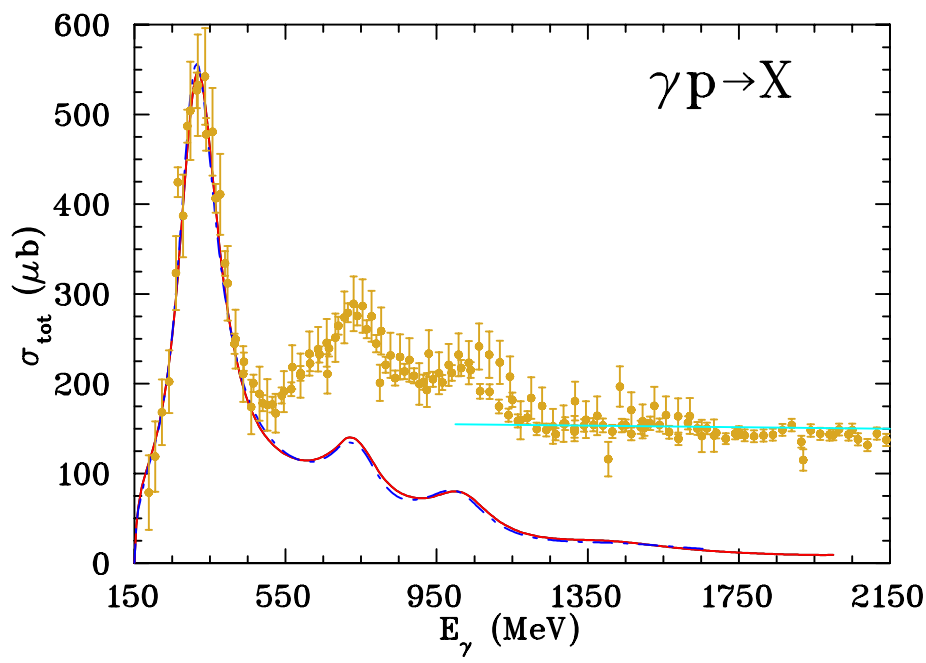}
    \includegraphics[width=0.45\textwidth,keepaspectratio]{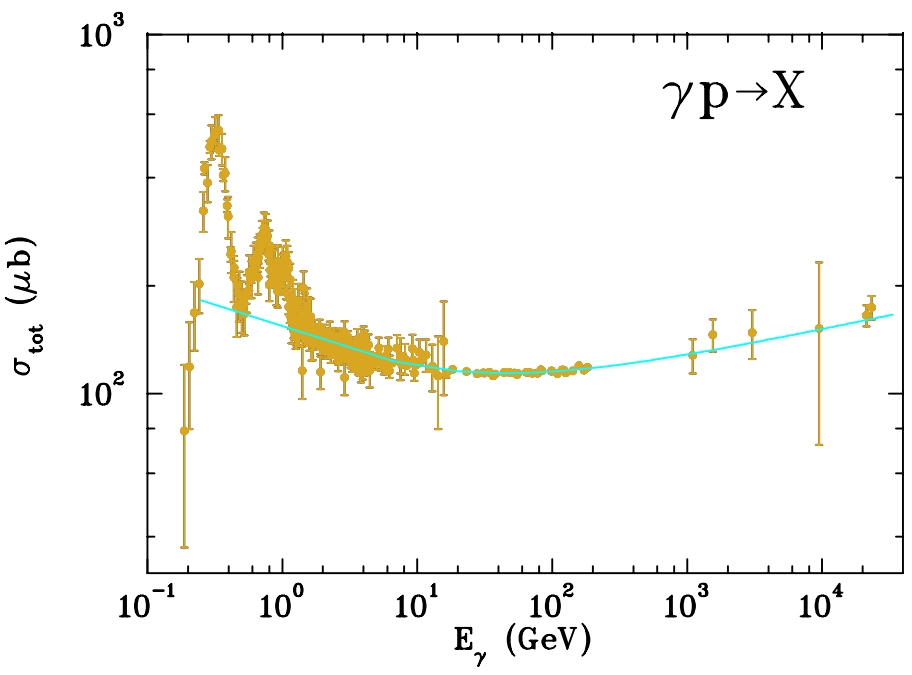}    
}
\caption{Top panel: Single-pion photoproduction contributions to the total cross section for a proton target, obtained in the SAID MA19 (\cite{A2:2019yud}, red solid line) and MAID2007 (\cite{Drechsel:2007if}, blue dash-dotted) analyses. The amplitudes from SAID (MAID) have been integrated up to $E_\gamma = 2~\mathrm{GeV}$ ($1.67~\mathrm{GeV}$). The proton total photoabsorption data are from Ref.~\cite{had-xsec20}. The total uncertainty is the quadratic sum of the statistical and systematical uncertainties. Bottom panel: Total photoabsorption cross section off the proton target at high energies. The cyan solid curve shows the Regge best fit result~\cite{Bass:2018uon} (see Eq.~(\ref{eq:Reg})).}
\label{fig:sigtot}
\end{figure}

Early estimates of the GDH integral used phenomenological single-pion photoproduction amplitudes and crude estimates for two-pion production and for production of mesons other than pions. Particularly important are the two-pion production contributions, as they dominate the total photoabsorption cross section over much of the resonance region. To illustrate this point, the SAID~\cite{A2:2019yud} and MAID~\cite{Drechsel:2007if} single-pion contributions to the total photoabsorption cross section are plotted in Fig.~\ref{fig:sigtot} (top). 

Some of the two-pion channels have been measured separately, and would appear to give a significant contribution to the GDH sum rule over much of the resonance region~\cite{Arndt:2005wk}.  In addition, as mentioned above, $\sigma_\mathrm{tot}$ begins to increase at high energies (Fig.~\ref{fig:sigtot} (bottom)), and its energy dependence can be parameterized in Regge formalism as~\cite{Bass:2018uon}
\begin{eqnarray}
\sigma_\mathrm{tot} = \left( 67.7 \, s^{+0.0808} + 129 \, s^{-0.4545} \right)\,\mu\mathrm{b}\>, 
    \label{eq:Reg}
\end{eqnarray}
where $s = M(M+2E_\gamma)$, in units of $\mathrm{GeV}^2$ for the above formula, which is applicable up to $\sqrt{s}\approx 250~\mathrm{GeV}$.  The first term is associated with gluonic pomeron exchange, and the second term is associated with the isoscalar $\omega$ and isovector $\rho$ trajectories.

A quite different behavior is observed if the measured helicity-dependent total cross section $\Delta\sigma$ is compared to its single-pion contribution over the resonance region.
Direct experimental verification of the GDH sum rule are difficult because of the need to extend the measurements to sufficiently high $E_\gamma$ in spite of the $1/E_\gamma$ weighting~\cite{Ireland:2019uwn}, and to reliably cover not only single- and double-pion photoproduction but all possible photon-induced processes.  One should be aware that some photoabsorption determinations of the GDH and GGT integrals encompassed only a restricted energy range.

Some of the most recent determinations of the GDH sum rule are compiled in Table~\ref{tab:tbl1} including electroproduction data taken at very low photon virtuality, down to $Q^2\approx 0.02~\mathrm{GeV}^2$~\cite{CLAS:2021apd,CLAS:2017ozc}. The data were extrapolated to $Q^2=0$ using guidance from chiral effective field theory. 
While such an extrapolation increases the systematic uncertainty, it is mitigated by the fact that the electroproduction experiments access simpler processes (inclusive electron scattering that automatically sums all reaction channels), in contrast to photoproduction experiments in which each final state must be measured, with additional associated systematics.
This results in electroproduction measurements of GDH being competitive in accuracy with respect to those involving photoproduction, as can be seen in Table~\ref{tab:tbl1}.
Sum rule determinations, involving integrals weighted by higher powers of $E_\gamma$, are listed for the Baldin sum rule in Table~\ref{tab:tbl2} and the GGT sum rule in Table~\ref{tab:tbl3}. We extrapolated to $Q^2=0$ the latest JLab electroproduction data on $\gamma_0^p$~\cite{CLAS:2021apd} and $\gamma_0^n$ (from $^3$He~\cite{E97-110:2021mxm} or deuteron~\cite{CLAS:2017ozc} targets) using chiral effective field theory calculations \cite{Bernard13}. The latter include state-of-the-art $\Delta$-isobar 
corrections and agree better with the lowest $Q^2$ data on $\gamma_0$ from~\cite{CLAS:2021apd,CLAS:2017ozc,E97-110:2021mxm} than other state-of-the-art calculations~\cite{Alarcon:2020icz}. For proton data, we extrapolated the weighted averaged of the three lowest-$Q^2$ data points ($\langle Q^2 \rangle=0.0145~\mathrm{GeV}^2$). We did likewise for the deuteron data ($\langle Q^2 \rangle=0.025~\mathrm{GeV}^2$) and extracted the neutron information from the proton and deuteron results at $Q^2=0$. For the neutron data obtained from $^3$He~\cite{E97-110:2021mxm} we used the lowest $Q^2$ point ($Q^2=0.035~\mathrm{GeV}^2$) for the extrapolation.

\begin{table}[htb!]
\caption{Compilation of GDH sum rule determinations for proton and neutron targets, in units of $\mu$b, in order of publication year (older to newer, post-2000 results only). The top part of the table gives experimental or phenomenological results, the middle part presents the Regge theoretical prediction, and the lower part presents results for the right hand side of the sum rule's definition (Eq.~(\ref{eq:GDH})), based on the anomalous magnetic moment. The CLAS EG4 data~\cite{CLAS:2021apd,CLAS:2017ozc} were taken at very low pion electroproduction $Q^2$, down to $\approx 0.01~\mathrm{GeV}^2$, and were extrapolated using the guidance of chiral effective field theory to $Q^2=0$ as described in Refs.~\cite{CLAS:2021apd,CLAS:2017ozc}.
	\label{tab:tbl1}}
\begin{center}
\begin{tabular}{|l|l|l|}
\hline
   Proton          & Neutron          & Lab/Ref   \\
\hline
 $168$             & $120$            & MAID~\cite{Kamalov:2000en} \\
 $176\pm 8$        &                  & GDH/A2~\cite{GDH:2000tuw} \\ 
 $226\pm 5\pm 12$  & $187\pm 8\pm 10$ & GDH/A2~\cite{GDH:2001zzk} \\
 $187$             & $137$            & SAID~\cite{Arndt:2002xv} \\
 $212\pm 6\pm 16$  &                  & Review~\cite{Helbing:2006zp} \\
 $254\pm 5\pm 12$  &                  & GDH~\cite{Dutz:2004zz} \\
 $168.8$           & $133.1$          & MAID~\cite{Drechsel:2007if} \\
 $204.5\pm 21.4$   &                  & Phenom~\cite{Gryniuk:2016gnm} \\
 $203\pm 11 $      & $235\pm 18$      & CLAS~\cite{CLAS:2021apd,CLAS:2017ozc} \\
\hline
 $211\pm 13$       &                  & Regge~\cite{Bass:2018uon} \\
 \hline
 $204.784482(35)$  & $232.25159(13)$  & PDG~\cite{ParticleDataGroup:2020ssz}\\
\hline
\end{tabular}
\end{center}
\end{table}

\begin{table}[htb!]
\caption{Compilation of Baldin sum rule determinations for proton and neutron targets, in units of $10^{-4}$~fm$^3$. The top section shows experimental and phenomenological determinations of the integral of Eq.~(\ref{eq:Baldin}), the middle section shows determinations of $\alpha+\beta$, while the bottom shows the PDG's evaluation of $\alpha+\beta$ to which we make comparisons.
	\label{tab:tbl2}}
\begin{center}
\begin{tabular}{|l|l|l|}
\hline
   Proton                         & Neutron       & Lab/Ref   \\
\hline
 $11.6$                           & $13.5$        & MAID~\cite{Kamalov:2000en} \\
 $13.25\pm 0.86_{-0.58}^{+0.23}$  &               & LEGS~\cite{Blanpied:2001ae} \\
 $11.5$                           & $12.9$        & SAID~\cite{Arndt:2002xv} \\
                                  & $15.2\pm 3.9$ & A2~\cite{Kossert:2003zf} \\
 $11.53$                          & $13.18$       & MAID~\cite{Drechsel:2007if}\\
\hline
 $14.0\pm 0.3$                    & $15.2\pm 0.5$ & Phenom~\cite{Levchuk:1999zy} \\
                                  & $15.2$        & SAL~\cite{Kolb:2000ix} \\
 $13.1\pm 0.9\pm 1.0$             &               & A2~\cite{deLeon:2001dnx} \\
 $14.0\pm 1.3\pm 0.6$             &               & FIAN~\cite{Baranov:2001jv} \\
 $16.4\pm 3.6$                    & $15.3\pm 5.4$ & MAX-lab~\cite{Lundin:2002jy} \\
 $15.50\pm 1.64$                  &               & HBChPT~\cite{Beane:2002wn} \\
                                  & $16.0\pm 2.8$ & HBChPT~\cite{Hildebrandt:2004hh} \\
 $13.9\pm 0.8$                    & $15.2\pm 2.5$ & Review~\cite{Schumacher:2005an} \\
 $15.1\pm 1.0$                    & $18.3\pm 3.1$ & N3LO-BChPT~\cite{Lensky:2009uv} \\
$15.5\pm 1.0 \pm 0.8$  &          & EFT~\cite{McGovern:2012ew} \\
                                  & $13.3\pm 2.8 \pm 1.6$& EFT~\cite{Griesshammer:2015ahu} \\
                                  & $15.2\pm 2.1$ & MAX-Lab~\cite{COMPTONMAX-lab:2014cve} \\
 $14.0\pm 0.2$                    &               & Phenom~\cite{Gryniuk:2015eza} \\
 $13.4\pm 1.4$                    &               & DR~\cite{Pasquini:2019nnx} \\
 $14.13\pm 0.26\pm 0.76$          &               & A2~\cite{A2:2021rcs} \\
 \hline
 $14.2\pm 0.5$                    & $15.5\pm 1.6$ & PDG~\cite{ParticleDataGroup:2020ssz} \\
\hline
\end{tabular}
\end{center}
\end{table}

\begin{table*}[htb!]
\caption{Compilation of GGT sum rule estimates for proton and neutron targets, in units of $10^{-4}$~fm$^4$. The top section lists experimental and phenomenological determinations, the middle section lists theoretical predictions, while the bottom shows the experimental average to which we compare.  The latest data on $\gamma_0$ for the proton~\cite{CLAS:2021apd} and the neutron (using $^3$He~\cite{E97-110:2021mxm} and deuteron~\cite{CLAS:2021apd} targets) have been obtained in electroproduction experiments at JLab. The data presented here are extrapolated to $Q^2=0$ using chiral effective field theory; see main text for details. For these data, the first listed uncertainty is statistical, the second is the experimental systematics, and the third is the extrapolation uncertainty.
	\label{tab:tbl3}}
\begin{center}
\begin{tabular}{|l|l|l|}
\hline
   Proton                 & Neutron & Lab/Ref   \\
\hline
 $-1.68\pm 0.10$           &                      & GDH/A2~\cite{GDH:2000tuw} \\
 $-0.68$                   & $-0.14$              & MAID~\cite{Kamalov:2000en} \\
 $-1.55\pm 0.15\pm 0.03$   &                      & LEGS~\cite{Blanpied:2001ae} \\
 $-1.87\pm 0.08\pm 0.10$   &                      & GDH/A2~\cite{GDH:2001zzk}     \\
 $-0.85$                   & $-0.08$              & SAID~\cite{Arndt:2002xv} \\
 $-0.72$                   & $-0.14$              & MAID~\cite{Drechsel:2007if} \\
 $-0.929\pm 0.105$         &                      & Phenom~\cite{Gryniuk:2016gnm} \\
 $-1.404\pm 0.131\pm 0.169\pm0.401$ & $-1.21\pm 0.74\pm 0.77\pm0.40$ & CLAS~\cite{CLAS:2021apd,CLAS:2017ozc}\\
                           & $-2.04\pm 0.13\pm 0.37\pm0.40$ & JLab Hall~A~\cite{E97-110:2021mxm} \\
\hline
 $-0.90$                   &                      & K-matrix~\cite{Kondratyuk:2001ys} \\
 $-2.6\pm 1.9$             & $+0.5\pm 1.0$        & HBChPT~\cite{Gasparyan10,Gasparyan11} \\
 $-1.74\pm 0.40$           & $-0.77\pm 0.40$      & BChPT~\cite{Bernard13} \\
 $-0.9\pm 1.4$             & $+0.03\pm 1.4$       & N3LO-BChPT~\cite{Lensky:2015awa} \\
 $-2.5\pm 0.4\pm 0.3$      & $-1.9\pm 0.4\pm 0.4$ & BChPT($+\Delta$)~\cite{Thurmann:2020mog} \\
\hline
$-1.40 \pm 0.45$ & $-1.88 \pm 0.50$ & Expt.~avg.~\cite{CLAS:2021apd,CLAS:2017ozc,E97-110:2021mxm}\\
\hline
\end{tabular}
\end{center}
\end{table*}

\section{Evaluations in the SAID framework}

In this study, we consider experimental input restricted to single-pion photoproduction. Although clearly incomplete, this contribution is expected to give a dominant contribution to the GDH, Baldin, and GGT sum rules. To evaluate this single-pion component (see Ref.~\cite{Arndt05} for our earlier attempt), phenomenological amplitudes obtained in PWAs of the SAID and MAID groups have been used.  Amplitudes from SAID (MAID) have been integrated up to $E_\gamma = 2~\mathrm{GeV}$ ($1.67~\mathrm{GeV}$) in the laboratory photon energy.

\begin{figure*}[hbtp]
\centering
{
    \includegraphics[width=0.8\textwidth,angle=0,keepaspectratio]{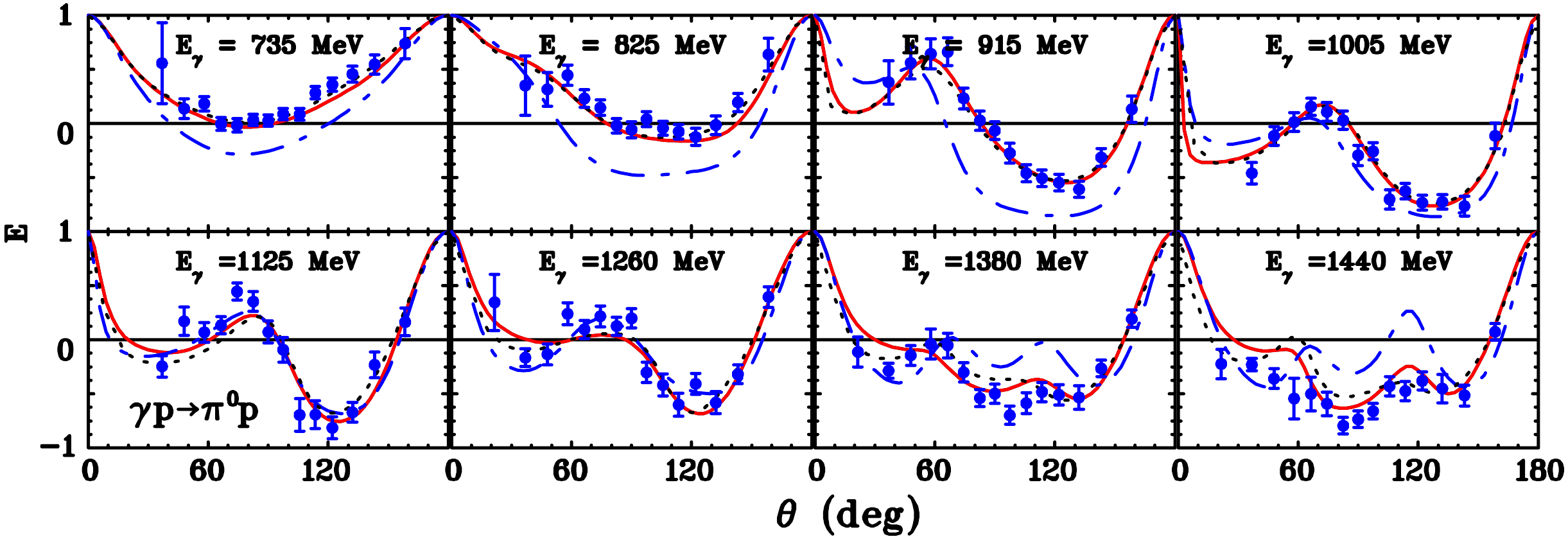}
    \includegraphics[width=0.8\textwidth,angle=0,keepaspectratio]{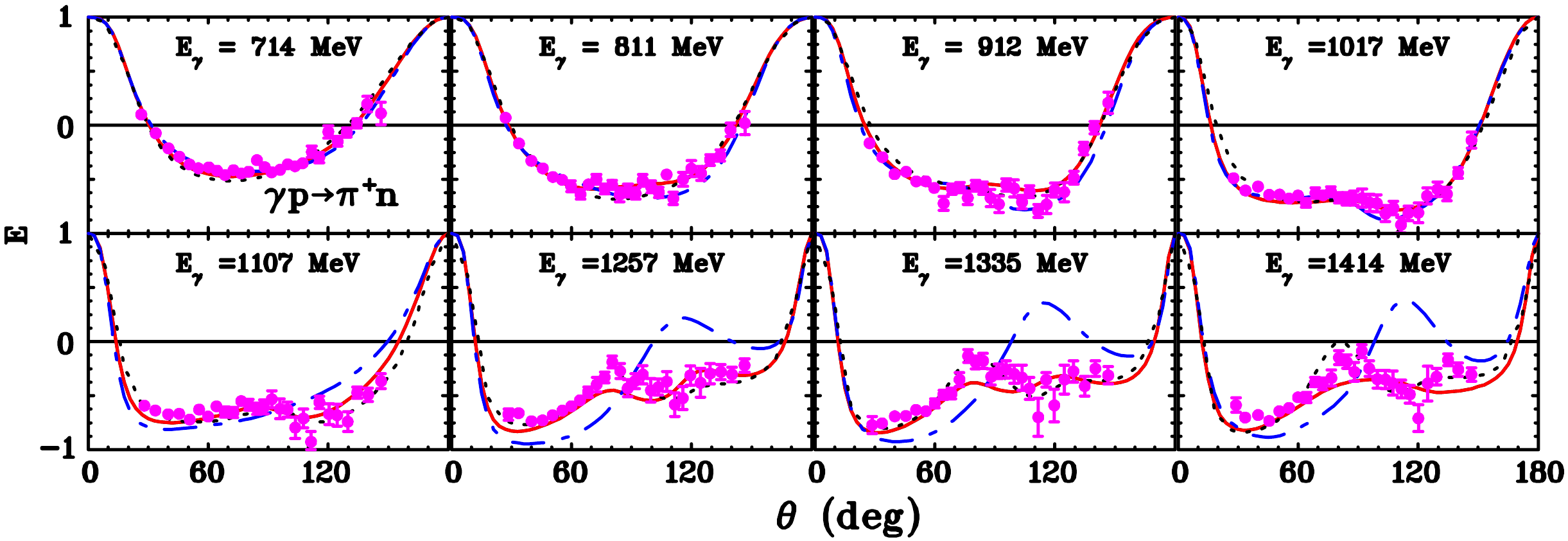}
    \includegraphics[width=0.8\textwidth,angle=0,keepaspectratio]{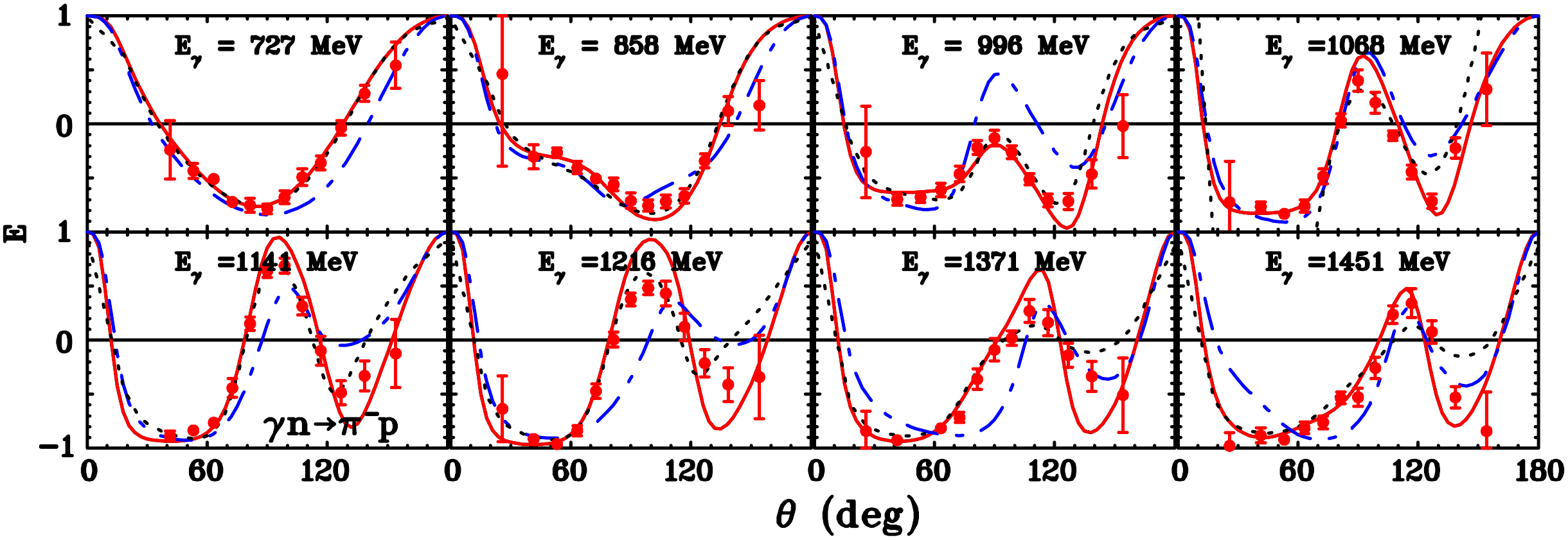}
    \includegraphics[width=0.8\textwidth,angle=0,keepaspectratio]{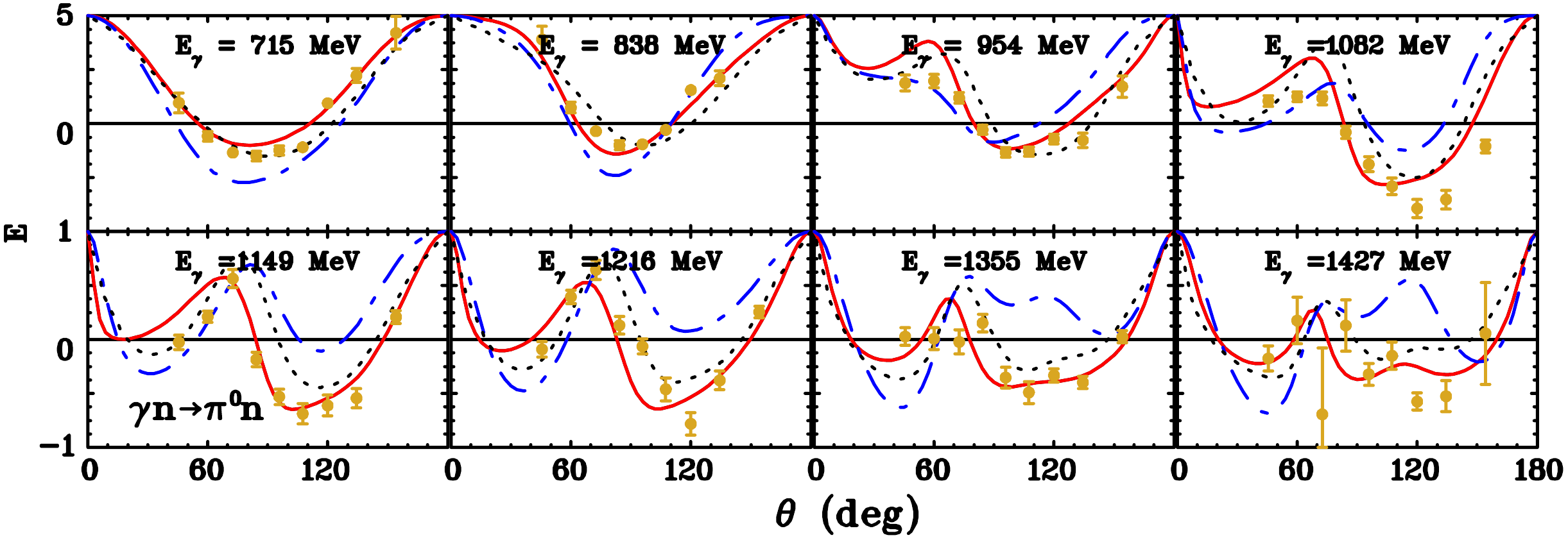}
}
\caption{Samples of double-polarized observable $\mathbb E$ for $\vec{\gamma}\vec{N}\to\pi N$ in the range of $E_\gamma = 700 - 1450~\mathrm{MeV}$. The red solid (blue dash-dotted) lines show the SAID MA19~\protect\cite{A2:2019yud} (MAID2007~\protect\cite{Drechsel:2007if}) solution,
while the Bonn-Gatchina solution BG2019 (which is an ongoing effort and represents an upgrade of the BG2014 solution reported in Ref.~\cite{BG2016}) is shown by black dotted lines. The experimental data for $\vec{\gamma}\vec{p}\to\pi^0p$ are from CBELSA/TAPS~\protect\cite{CBELSATAPS:2013btn}; for $\vec{\gamma}\vec{p}\to\pi^+n$ they are from CLAS/FROST~\protect\cite{CLAS:2015ykk}; for $\vec{\gamma}\vec{n}\to\pi^-p$ they are from A2~\protect\cite{Dieterle:2017myg}, and for $\vec{\gamma}\vec{n}\to\pi^0n$ from CLAS/HD-Ice~\protect\cite{CLAS:2017kua}.}  
\label{fig:E}
\end{figure*}

In Fig.~\ref{fig:E}, the available data for $\mathbb E$ from the four possible pion photoproduction reactions ($\vec{\gamma} \vec{p}\to\pi^0p$, $\vec{\gamma} \vec{p}\to\pi^+n$, $\vec{\gamma}\vec{n}\to\pi^0n$ and $\vec{\gamma}\vec{n}\to\pi^-p$) are compared over the photon energy range $E_\gamma = 700 - 1450~\mathrm{MeV}$. Here we also display an updated version of the Bonn-Gatchina analysis~\cite{BG2016} to show the overall consistency of the most recent fits. 

The difference of cross-sections for helicity states $3/2$ and $1/2$, that is, $\Delta(d\sigma/d\Omega) = (d\sigma_{3/2}/d\Omega - d\sigma_{1/2}/d\Omega)$, for $\vec{\gamma}\vec{N}\to\pi N$, is given in terms of helicity amplitudes

\begin{eqnarray*}
	\frac{d\sigma_{3/2}}{d\Omega} &=& \frac{q}{k}\left(|H_1|^2 + |H_3|^2\right) \>, \\
	\frac{d\sigma_{1/2}}{d\Omega} &=& \frac{q}{k}\left(|H_2|^2 + |H_4|^2\right) \>,
\end{eqnarray*}
where $q$ and $k$ are the pion and photon center-of-mass momenta, respectively. Their sum and difference can then be used to construct the beam-target polarization quantity $\mathbb E$~\cite{Barker:1975bp,Workman:1991xs} as
\begin{equation}
	\mathbb{E}~=~\frac{|H_2|^2 + |H_4|^2 - |H_1|^2 - |H_3|^2} {|H_2|^2 + |H_4|^2 + |H_1|^2 + |H_3|^2} \>.
    \label{eq:E}
\end{equation}
The recent SAID PWA solution MA19~\cite{A2:2019yud} uses all available $\mathbb E$ 
measurements~\cite{CBELSATAPS:2013btn,CLAS:2015ykk,Dieterle:2017myg,CLAS:2017kua} in the fit. 

\begin{figure*}[hbtp]
\vspace{0.4cm}
\centering
{
    \includegraphics[width=0.45\textwidth,keepaspectratio]{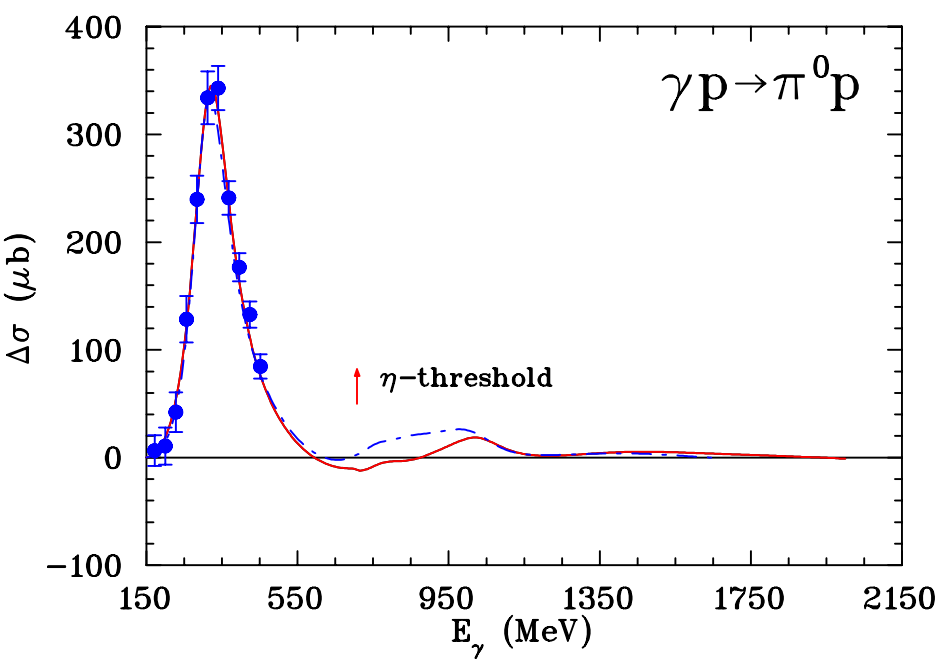}
    \includegraphics[width=0.45\textwidth,keepaspectratio]{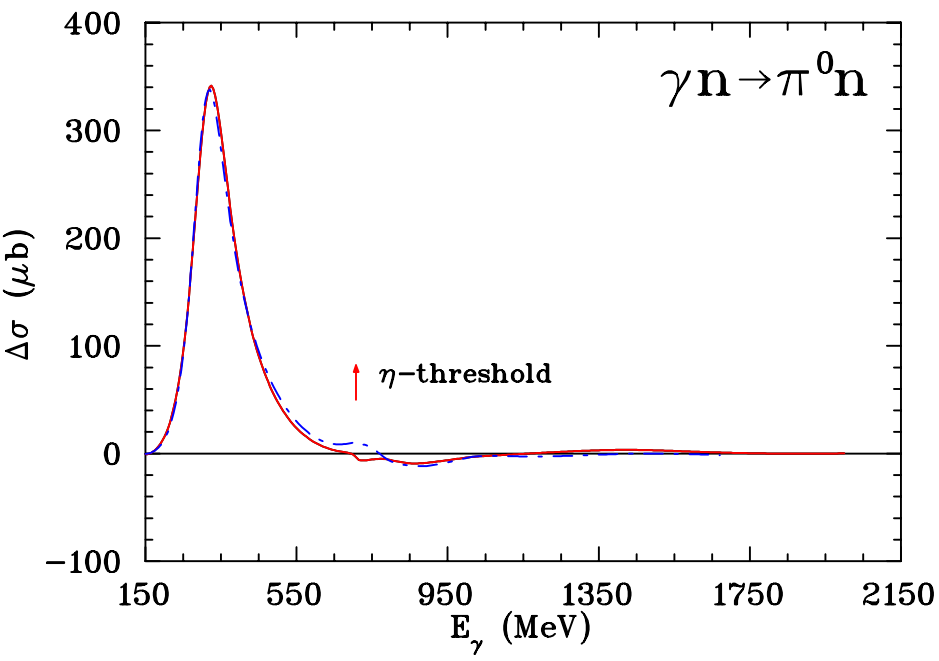}
}
\centering
{
    \includegraphics[width=0.45\textwidth,keepaspectratio]{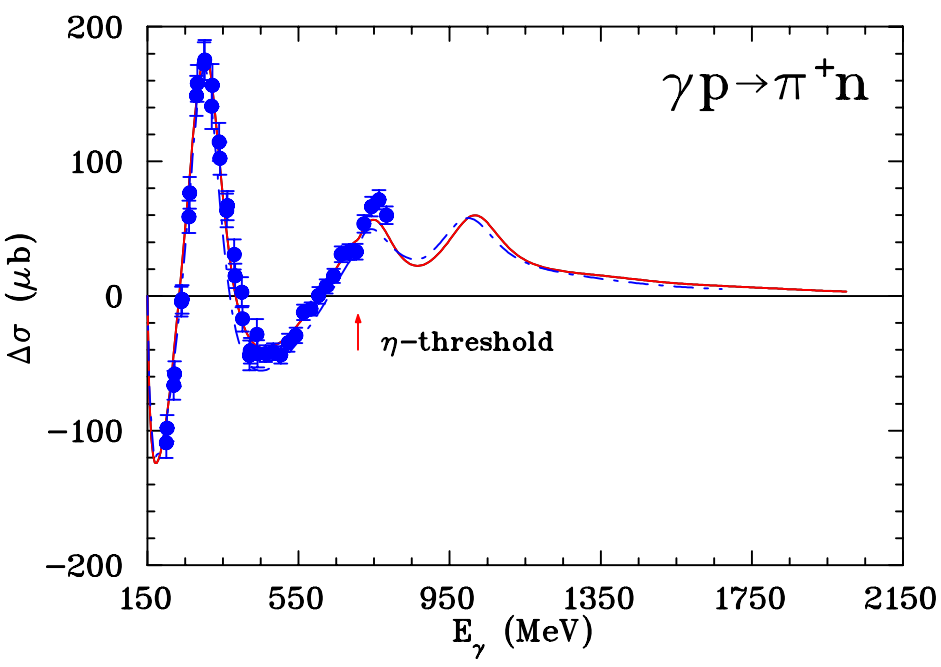}
    \includegraphics[width=0.45\textwidth,keepaspectratio]{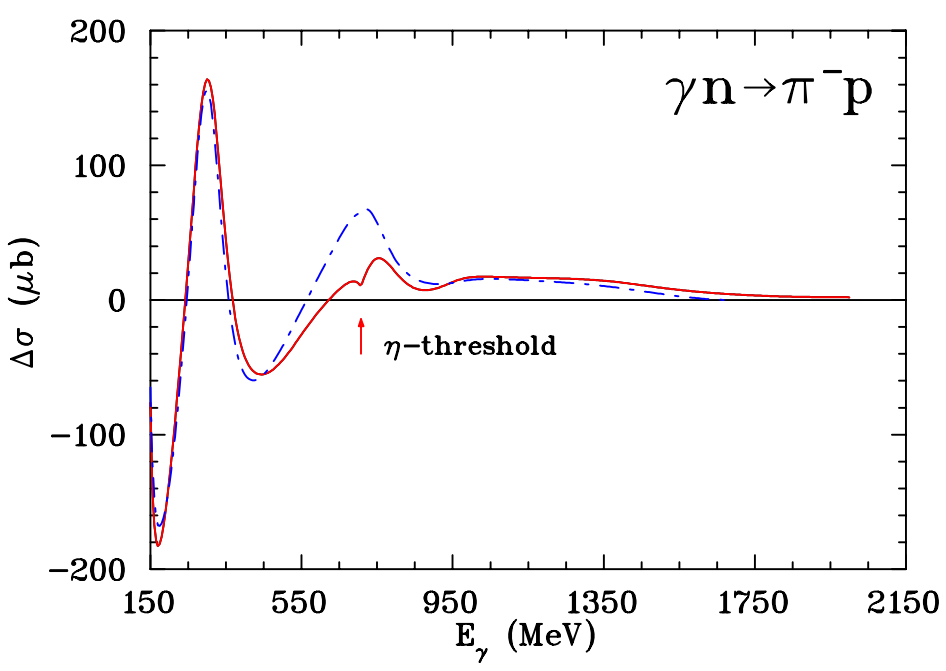}
}
\centering
{
    \includegraphics[width=0.45\textwidth,keepaspectratio]{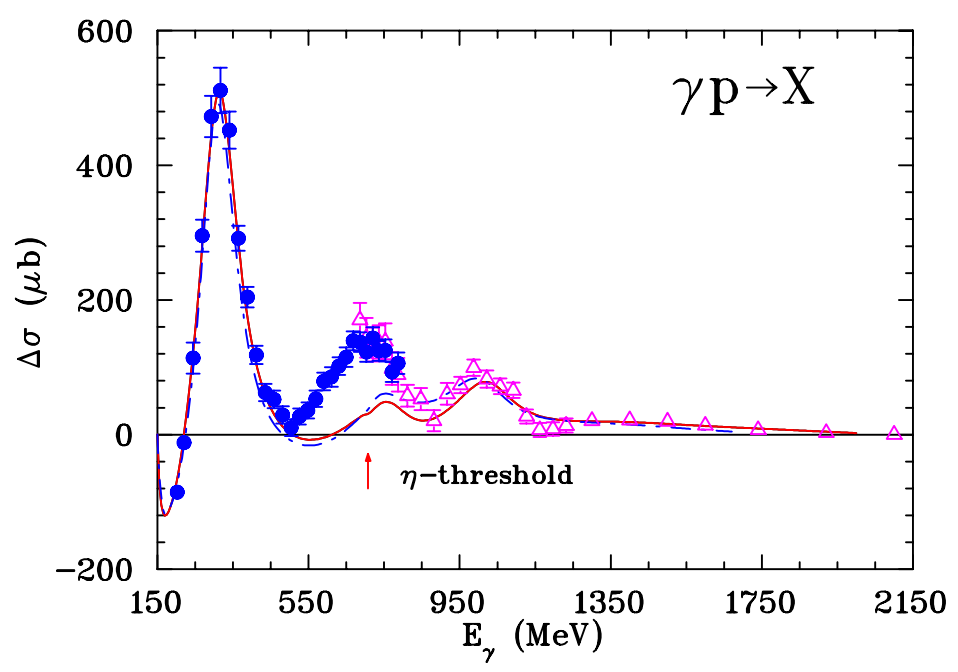}
    \includegraphics[width=0.45\textwidth,keepaspectratio]{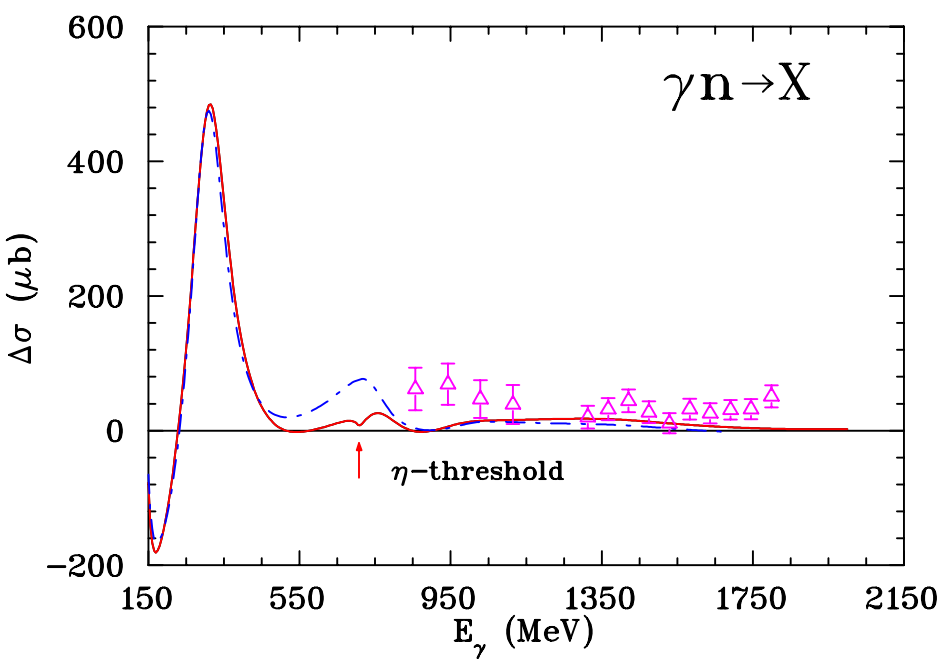}
}
\caption{Differences of total cross section for the helicity states $3/2$ and $1/2$, $\Delta\sigma$, for $\vec{\gamma}\vec{N}\to\pi N$. The red solid (blue dash-dotted) lines plot the SAID MA19~\protect\cite{A2:2019yud} (MAID2007~\protect\cite{Drechsel:2007if}, which is valid to $E_\gamma = 1670~\mathrm{MeV}$) solution. The experimental data for $\vec{\gamma}\vec{p}\to\pi^0p$ are from A2~\protect\cite{GDH:2004ydy,GDH:2000tuw}; for $\vec{\gamma}\vec{p}\to\pi^+n$ they are from A2~\protect\cite{GDH:2004ydy,Ahrens:2006gp,GDH:2000tuw}; for $\vec{\gamma}\vec{p}\to X$~\protect\cite{GDH:2001zzk,GDH:2003xhc} and for $\vec{\gamma}\vec{n}\to X$~\protect\cite{GDH:2005noz} they are from CBELSA. The red vertical arrow indicates the $\eta$-photoproduction threshold.}
\label{fig:deltasig}
\end{figure*}

The existing data on $\Delta\sigma$ are shown in Fig.~\ref{fig:deltasig} along with the SAID and MAID phenomenological (single-pion) estimates.  In both of these PWAs, the second (and $\eta$ threshold) as well as the third resonance regions 
around $700~\mathrm{MeV}$ and $1~\mathrm{GeV}$, respectively, are very pronounced for charged pions in the final state, whereas they are weak in the case of neutral pions. In the $\Delta$-isobar region, the predictions for both cases are the same.  The $\Delta\sigma$ for $\vec{\gamma}\vec{p}\to\pi^0p$ and $\vec{\gamma}\vec{p}\to\pi^+n$, 
measured by the A2 Collaboration at MAMI and GDH Collaboration~\cite{GDH:2004ydy,Ahrens:2006gp,GDH:2000tuw}, is well described by both
SAID MA19 and MAID2007 solutions. The CBELSA~\cite{GDH:2001zzk,GDH:2003xhc} and MAMI~\cite{GDH:2004ydy,Ahrens:2006gp,GDH:2000tuw} experimental data for $\vec{\gamma}\vec{p}\to X$ are in good agreement but disagree with PWA predictions for $\vec{\gamma}\vec{p}\to\pi N$. Around the $\eta$-threshold, the deviation becomes more apparent. This is because the contributions of the double-pion and $\eta$-meson production are missing here. 

\begin{figure*}[ht]
\vspace{0.4cm}
\centering
{
    \includegraphics[width=0.38\textwidth,angle=90,keepaspectratio]{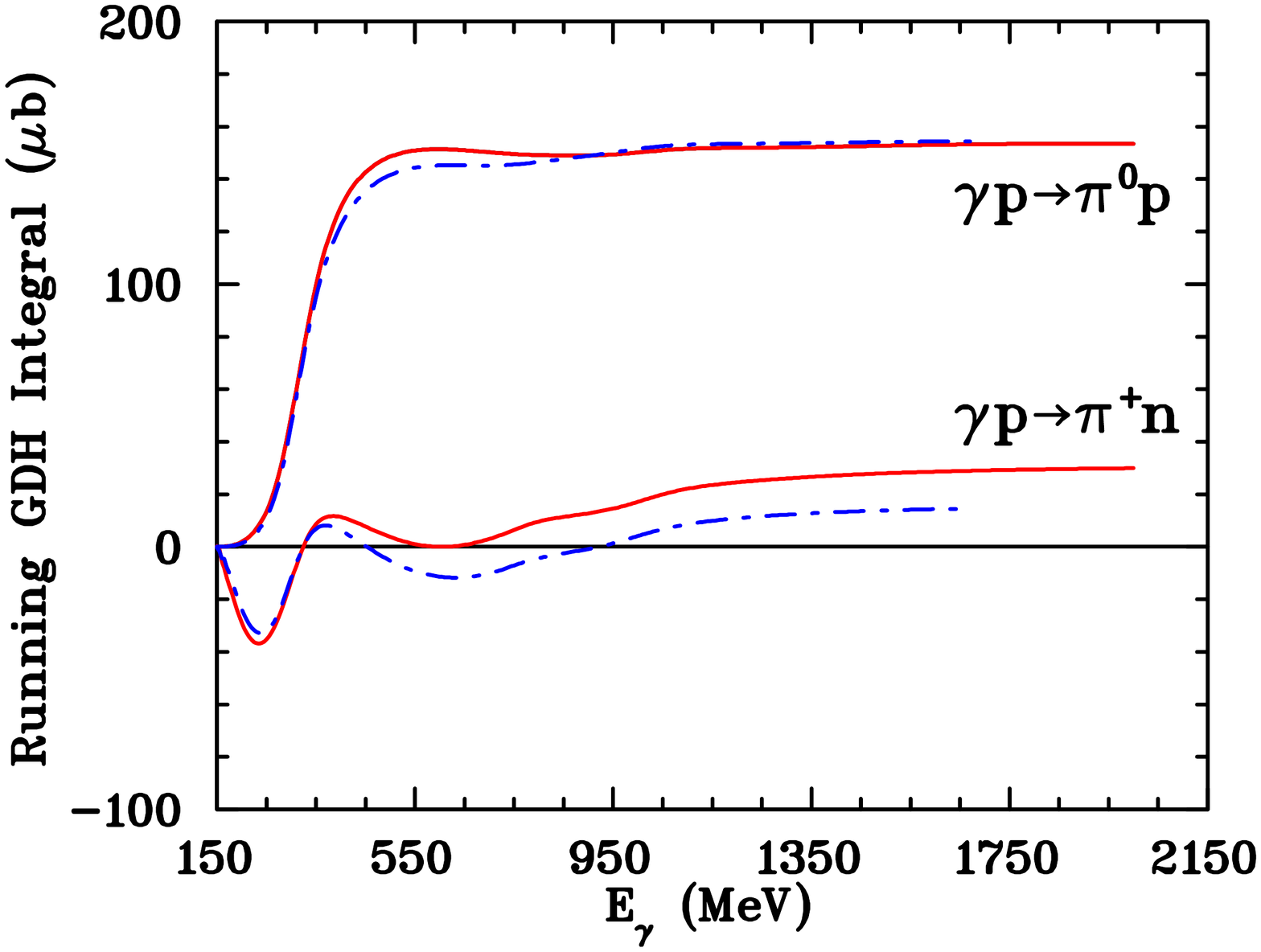}
    \includegraphics[width=0.38\textwidth,angle=90,keepaspectratio]{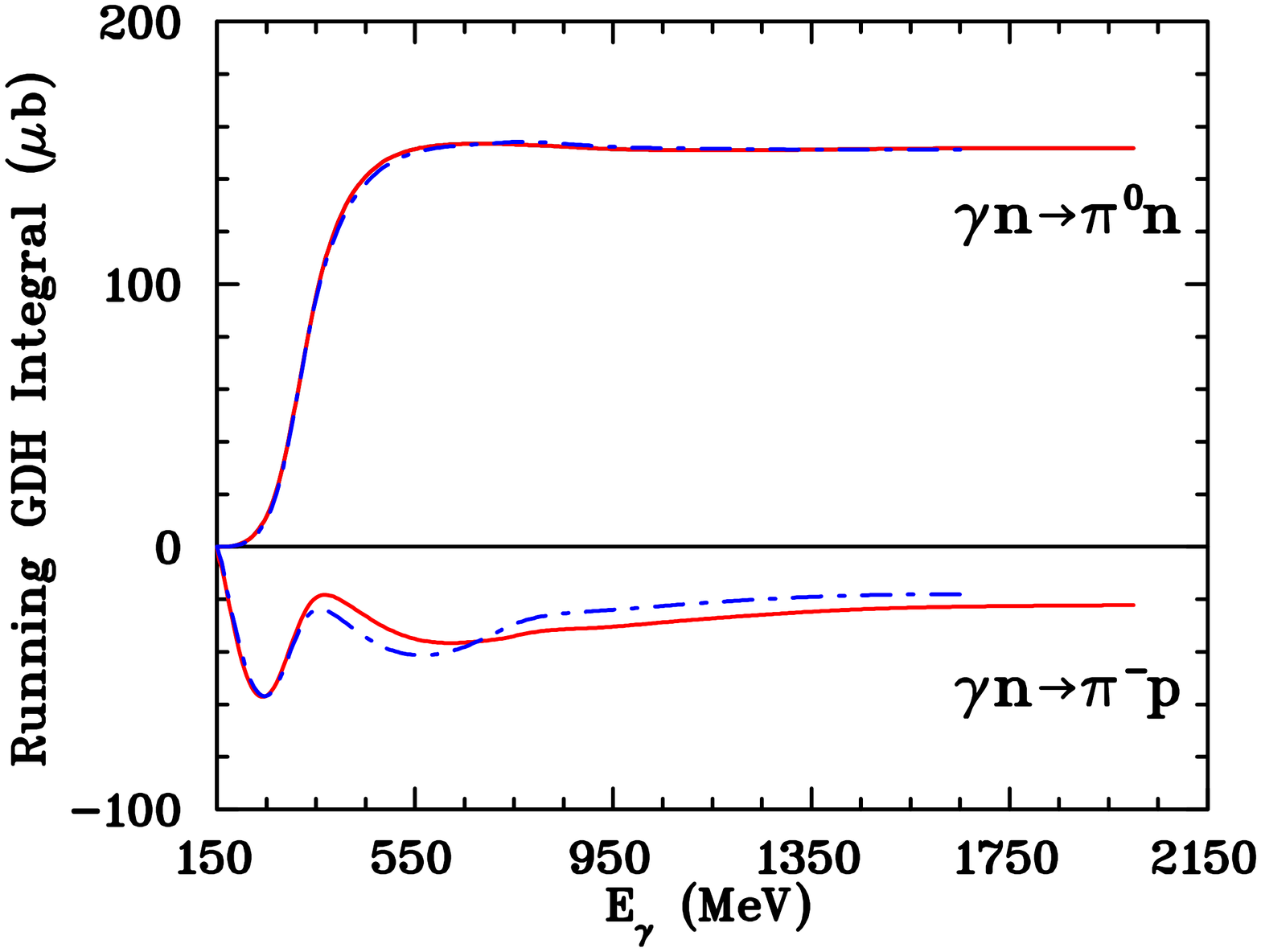}
}
\centering
{
    \includegraphics[width=0.38\textwidth,angle=90,keepaspectratio]{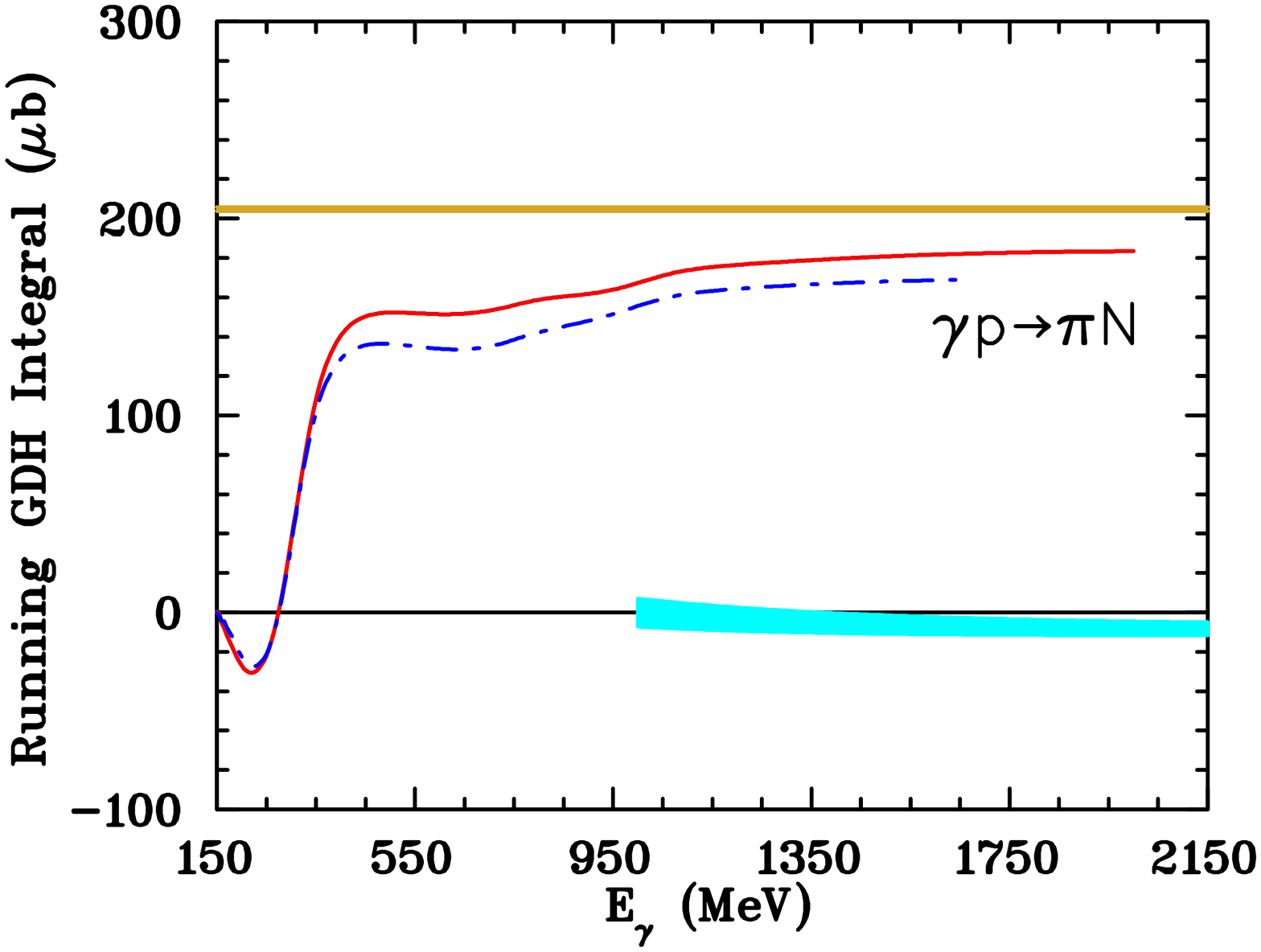}
    \includegraphics[width=0.38\textwidth,angle=90,keepaspectratio]{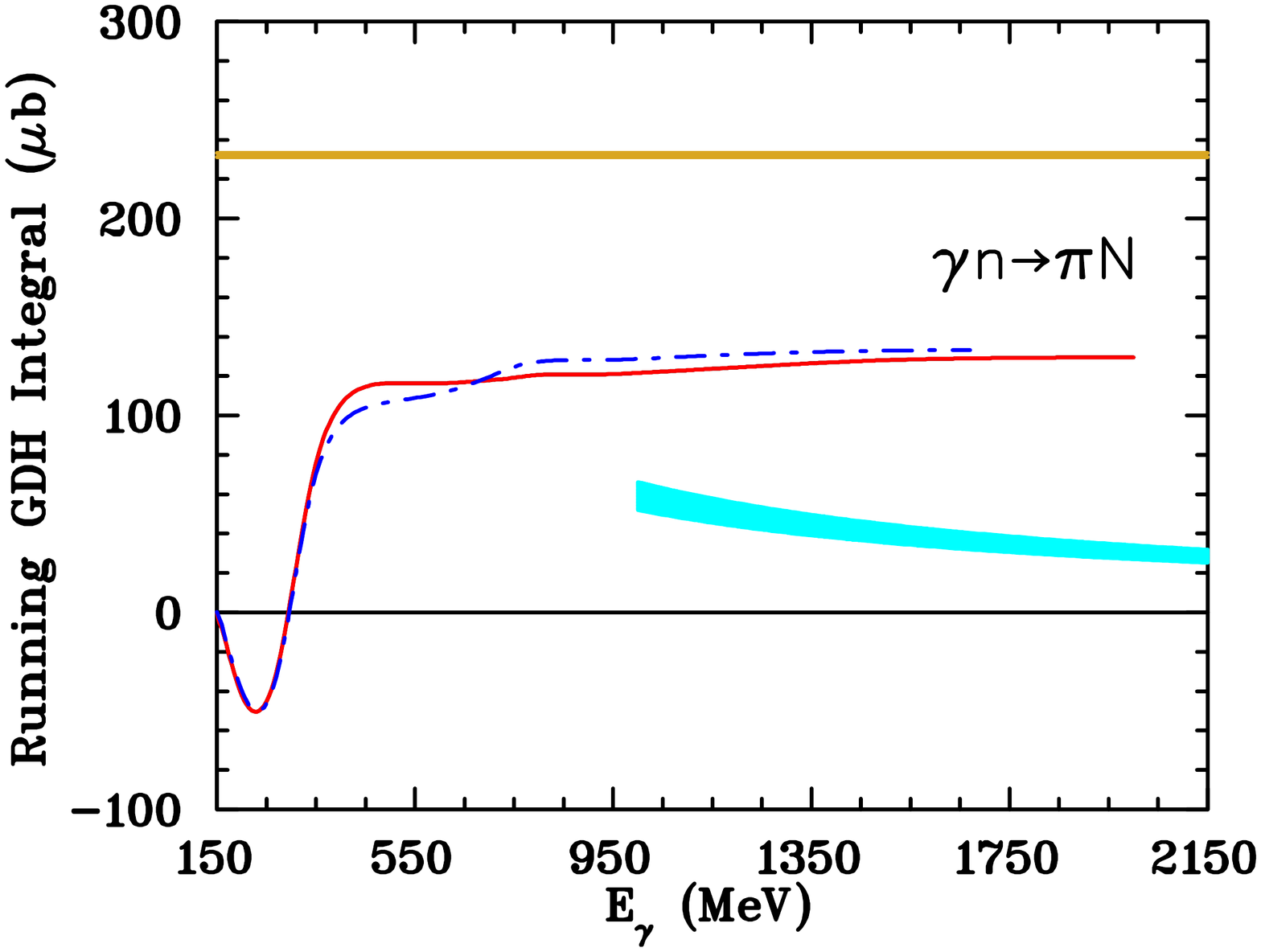}
}
\caption{Running GDH integral for proton (left) and neutron (right) targets. The red solid (blue dash-dotted) lines represent the SAID MA19~\protect\cite{A2:2019yud} (MAID2007~\protect\cite{Drechsel:2007if}) solution. The thick yellow horizontal lines in the bottom panels show the predicted GDH integral values calculated with the measured anomalous magnetic moments of the proton and neutron, see Eq.~(\ref{eq:GDH}); the uncertainties are too small to be visible on this plot (see Table~\ref{tab:tbl1}). The cyan bands show the Regge estimates of the complement to the running integral, that is, not from
threshold to the current photon energy, but rather from the current photon energy to infinity.}
\label{fig:GDHrun}
\end{figure*}

The running GDH integrals,
\begin{equation}
    \int_{E_\gamma^\mathrm{thr}}^{E_\gamma}\frac{\Delta\sigma(E_\gamma')}{E_\gamma'}\,dE_\gamma' \>,
\end{equation}
for the proton and neutron are shown in Fig.~\ref{fig:GDHrun}, where a comparison of the SAID with MAID single pion photoproduction results is also given.  Both the proton and neutron running integrals corresponding to single-pion photoproduction (bottom left and right panels, respectively) indicate convergence but remain lower than the predicted values (thick yellow lines in the bottom panels of Fig.~\ref{fig:GDHrun}).  In the proton case ($I_\mathrm{GDH} = 183.4~\mu$b), approximately 21~$\mu$b strength (which is about 10\% of the right-hand-side of Eq.~(\ref{eq:GDH})) is missing and can be attributed to non-single-pion processes, while the gap is much larger in the neutron case ($I_\mathrm{GDH} = 129.5~\mu$b), where $104~\mu$b are missing (about 44\% of the right-hand-side of Eq.~(\ref{eq:GDH})). In the bottom panels of Fig.~\ref{fig:GDHrun}, we also show the contributions to the GDH running integral evaluated in the Regge approach (see Sec.~\ref{sec:Regge}). These contributions (cyan bands) are not evaluated from threshold to the current energy, but rather from the current photon energy to infinity, and hence should be added to the respective SAID and MAID curves. Since Regge theory only becomes applicable at energies far above the resonance region, their dependence is shown only above $E_\gamma = 1~\mathrm{GeV}$.

\begin{figure*}[ht]
\vspace{0.4cm}
\centering
{
    \includegraphics[width=0.38\textwidth,angle=0,keepaspectratio]{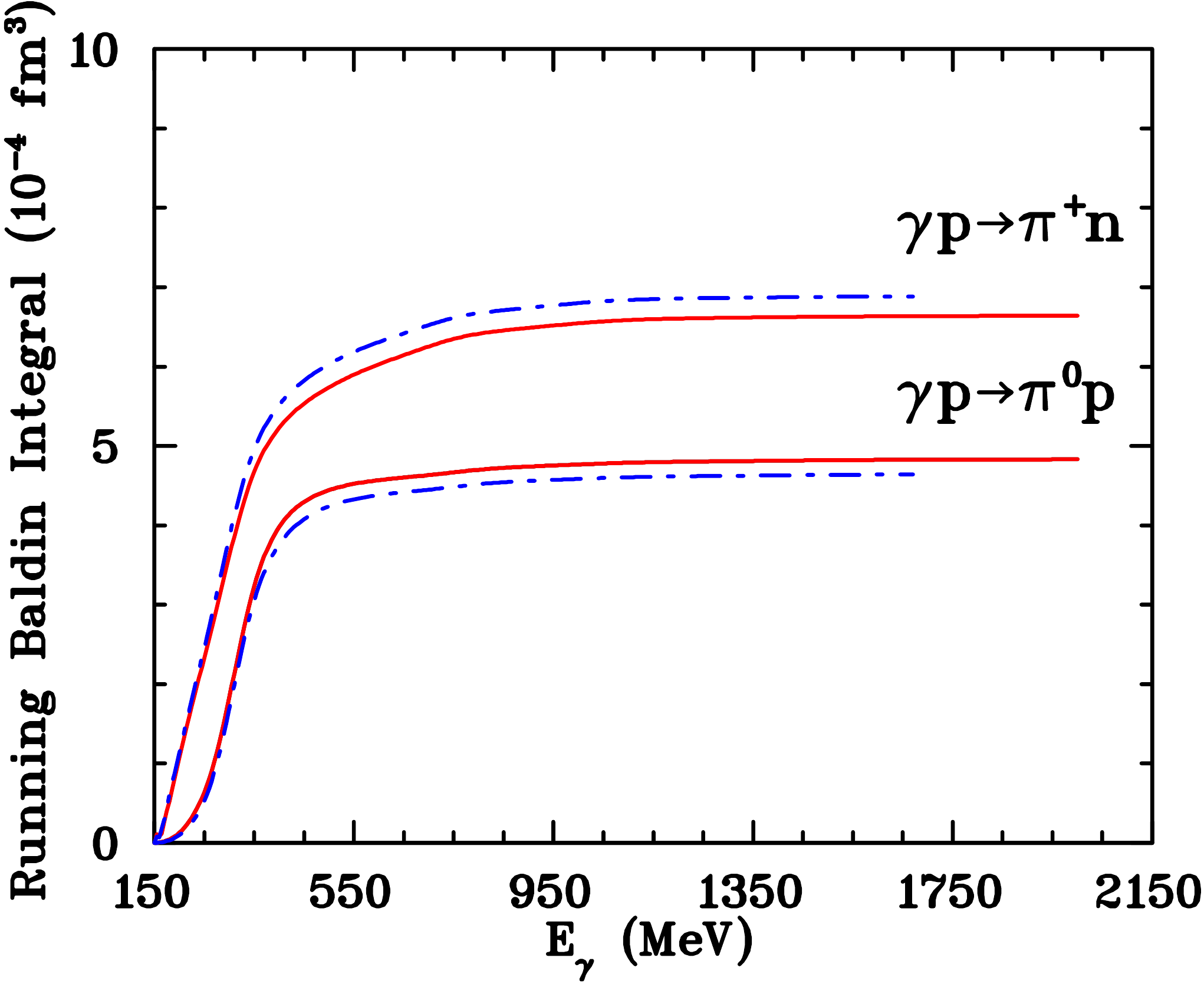}
    \includegraphics[width=0.38\textwidth,angle=0,keepaspectratio]{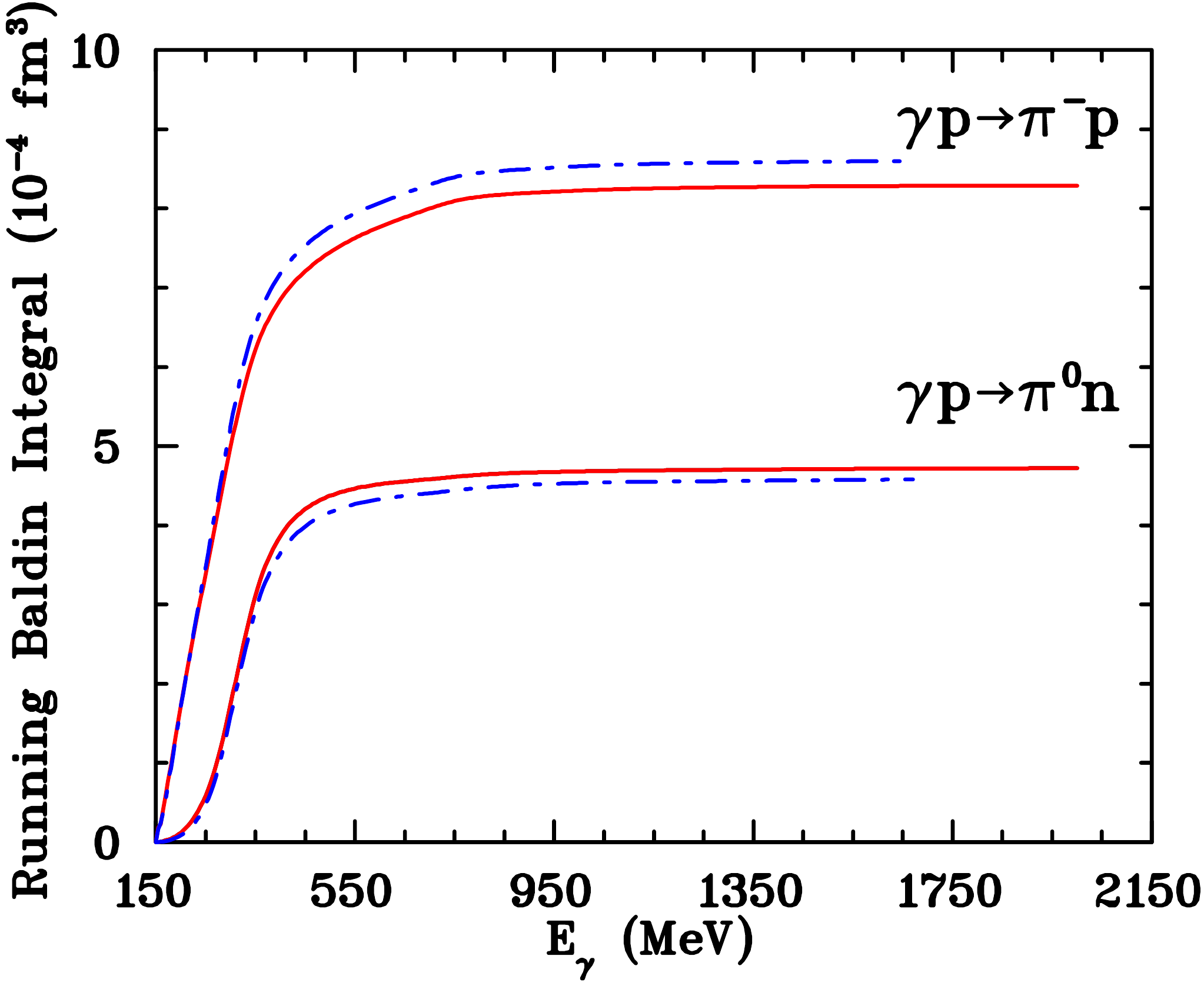}
}
\centering
{
    \includegraphics[width=0.38\textwidth, height=5.5cm]{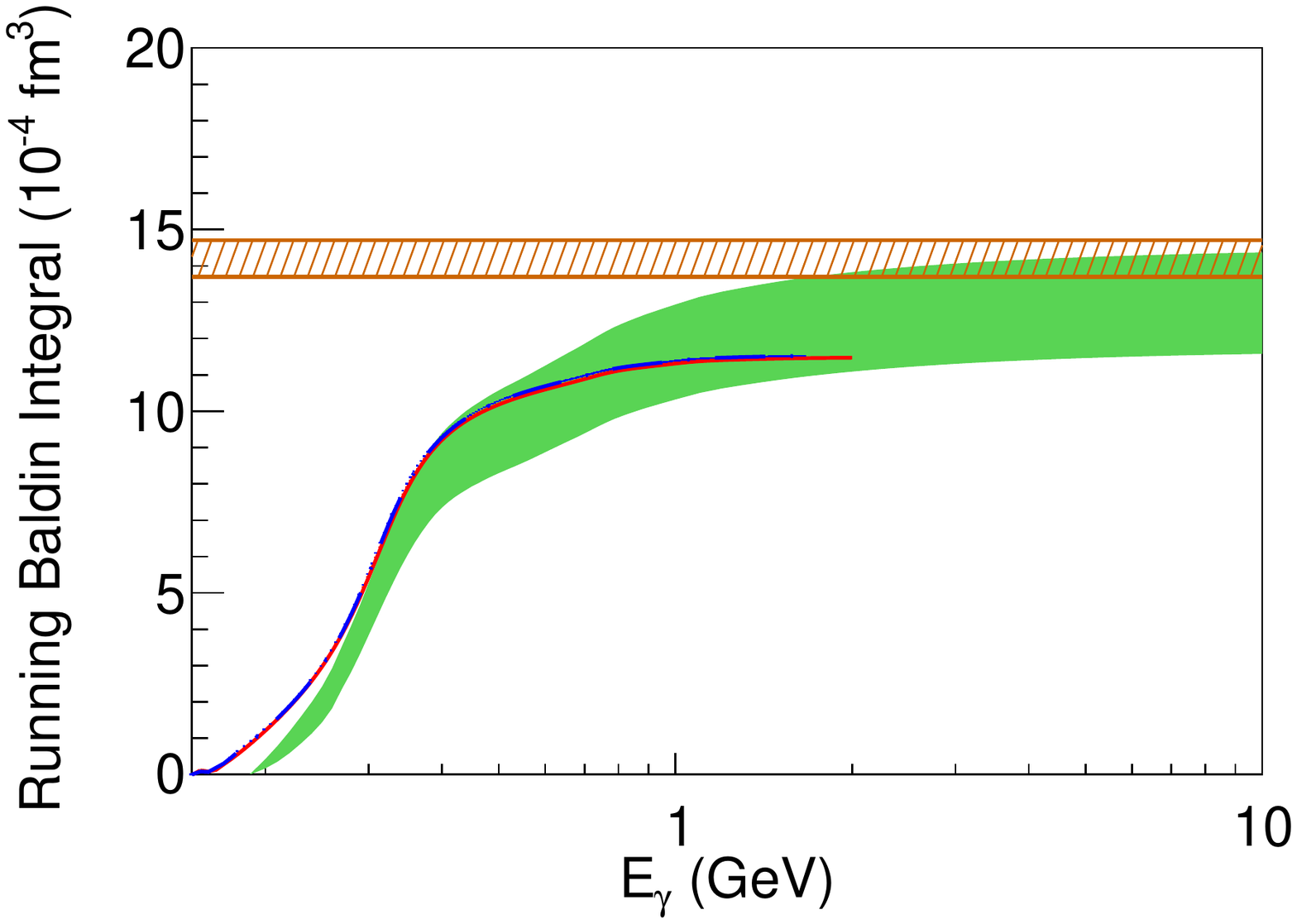}
    \includegraphics[width=0.38\textwidth,angle=0,keepaspectratio]{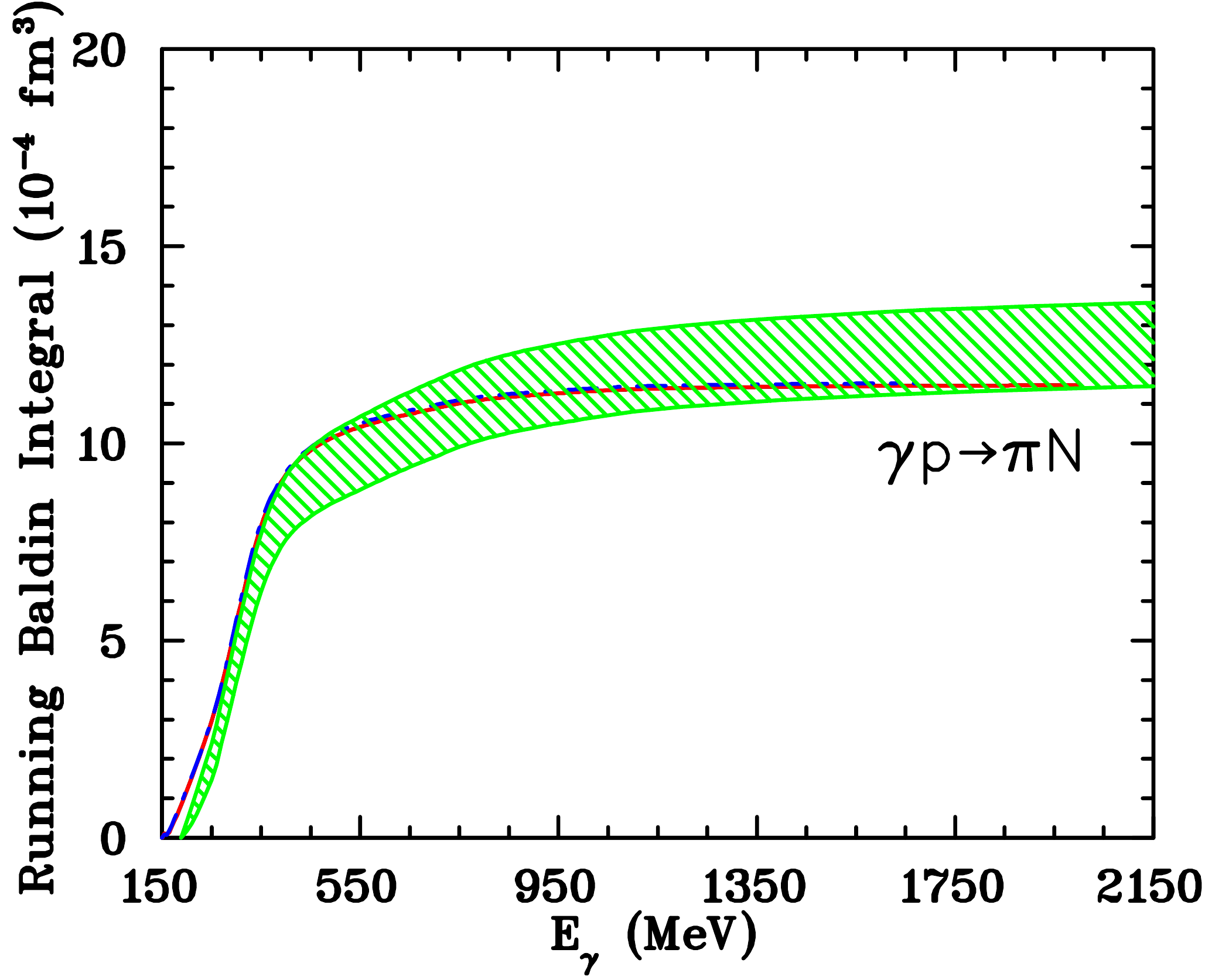}
}
\caption{Running Baldin integral for proton (left) and neutron (right) targets. The red solid (blue dash-dotted) lines represent the SAID MA19~\protect\cite{A2:2019yud} (MAID2007~\protect\cite{Drechsel:2007if}) solution. The yellow bands in the bottom panels show the PDG values obtained by averaging the data from various Compton scattering experiments~\cite{ParticleDataGroup:2020ssz} (see Table~\ref{tab:tbl2}).
The green band in the left bottom panel shows the running Baldin integral evaluated from total photoabsorption data on the proton target presented in Fig.~\ref{fig:sigtot}. The spread of the band represents statistical and systematical uncertainties combined in quadrature.}
\label{fig:Baldin}
\end{figure*}

The running Baldin integrals for the proton and neutron are shown in Fig.~\ref{fig:Baldin}, where a comparison of the SAID and MAID results is also given. In the proton case, the single-pion part gives $I_\mathrm{Baldin} = 11.47\times 10^{-4}~\mathrm{fm}^3$, hence the contribution from processes other than single-pion production (comparing Compton scattering results~\cite{ParticleDataGroup:2020ssz} and the phenomenological SAID PWA)  is  $2.7\times 10^{-4}~\mathrm{fm}^3$, which is about 19\% of the right-hand-side of Eq.~(\ref{eq:Baldin})). In contrast to the GDH running integral, the gap in the neutron case is consistent with the one found in the proton: the single-pion part for the neutron target gives $I_\mathrm{Baldin} = 13.01\times 10^{-4}~\mathrm{fm}^3$, leaving us with a missing strength of $2.5\times 10^{-4}~\mathrm{fm}^3$, which is about 16\% of the right-hand-side of Eq.~(\ref{eq:Baldin}).

Additionally, Fig.~\ref{fig:Baldin} (bottom left) shows the running Baldin integral evaluated from total photoabsorption data on the proton target presented in Fig.~\ref{fig:sigtot}. We fitted the 
data~\cite{had-xsec20} that spans photon energies from $188~\mathrm{MeV}$ to $23.3~\mathrm{TeV}$.  The evaluation yields $(13.1\pm 1.4)\times 10^{-4}~\mathrm{fm^3}$, which agrees well with the Compton estimate of $(14.2\pm 0.5)\times 10^{-4}~\mathrm{fm^3}$~\cite{ParticleDataGroup:2020ssz}.

\begin{figure*}[hbtp]
\vspace{0.4cm}
\centering
{
    \includegraphics[width=0.38\textwidth,angle=90,keepaspectratio]{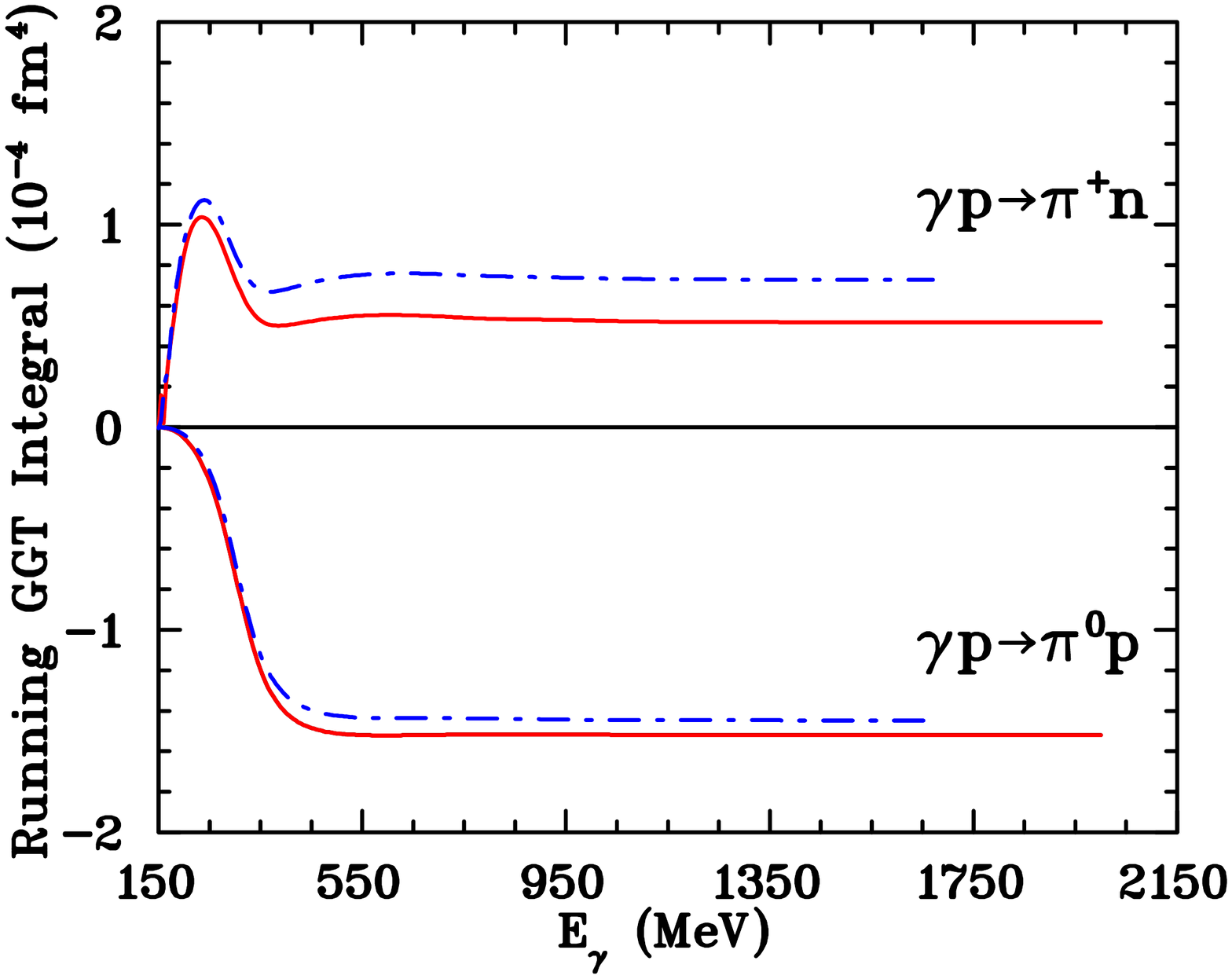}
    \includegraphics[width=0.38\textwidth,angle=90,keepaspectratio]{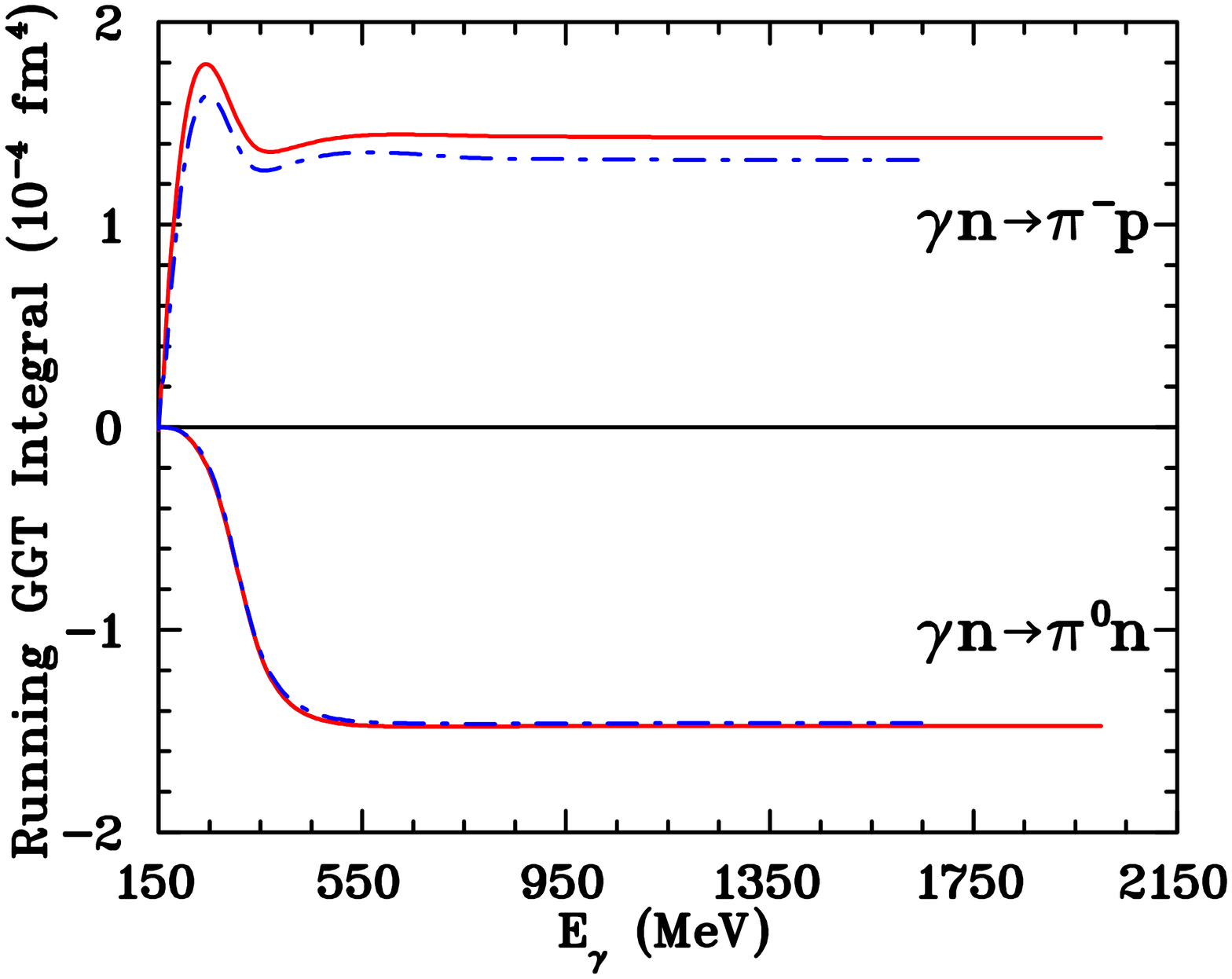}
}
\centering
{
    \includegraphics[width=0.38\textwidth,angle=90,keepaspectratio]{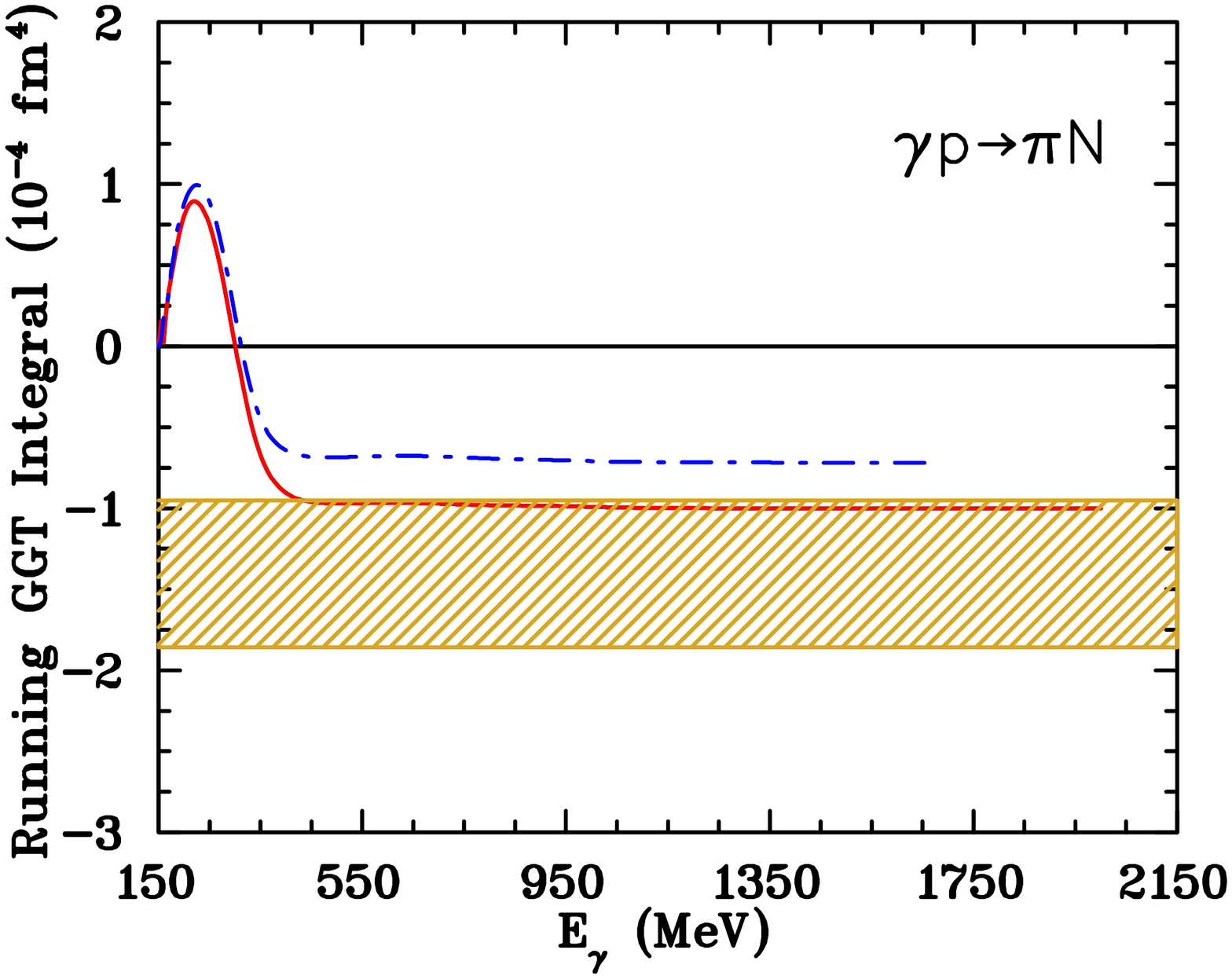}
    \includegraphics[width=0.38\textwidth,angle=90,keepaspectratio]{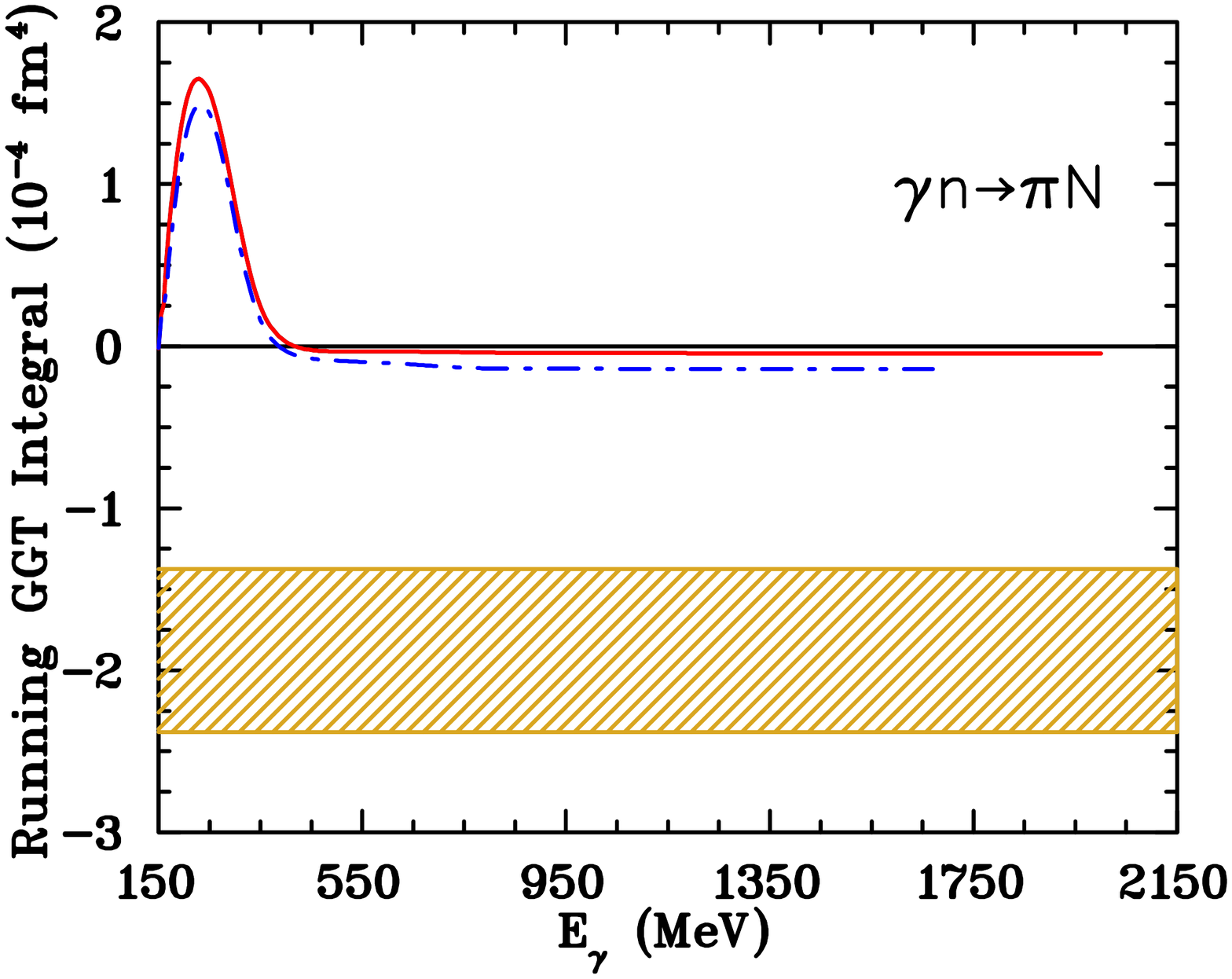}
}
\caption{ 
    Running GGT integral for proton (left) and neutron (right) targets. The red solid (blue dash-dotted) lines represent the SAID MA19~\protect\cite{A2:2019yud} (MAID2007~\protect\cite{Drechsel:2007if}) solution. The yellow bands in the bottom panels indicate the most recent values of $\gamma_0$ measured in pion electroproduction at JLab~\cite{CLAS:2021apd, CLAS:2017ozc,E97-110:2021mxm} and extrapolated to $Q^2=0$.
    \label{fig:GGT}}
\end{figure*}

The running integrals for the GGT sum rule for the proton and neutron are shown in Fig.~\ref{fig:GGT}, where a comparison of the SAID and MAID results is also given. In the SAID single-pion PWA, we obtain $\gamma_0 = -1.00\times 10^{-4}~\mathrm{fm}^4$ ($\gamma_0 = -0.04\times 10^{-4}~\mathrm{fm}^4$) for the proton (neutron).
While in the proton case the single-pion contribution of about $-1\times 10^{-4}\,\mathrm{fm}^4$ seems to point in the right direction, it is consistent with zero in the neutron case.
Also shown in Fig.~\ref{fig:GGT} are the latest measured values of $\gamma_0$ in pion electroproduction from JLab~\cite{CLAS:2021apd,CLAS:2017ozc,E97-110:2021mxm} extrapolated
to $Q^2=0$ (Table~\ref{tab:tbl3}). The width of the bands represents 
the uncertainties combined quadratically. The band for the neutron represents the weighted average of CLAS and JLab Hall~A E97--110 results.
According to the experimental values of the GGT integrals, the SAID single-pion production analysis yields conclusions similar to 
those for GDH: for the proton the SAID integral saturates near the value 
expected from the sum rule, while for the neutron, it misses a noticeable integral-strength. This is not surprising since the GDH and GGT integrands differ only by $1/E^2_\gamma$. 

\section{The GDH integrand in the Regge framework}
\label{sec:Regge}

As demonstrated above, the contributions to the GDH integral of processes beyond single-pion photoproduction are much smaller for the proton (where about $10\,\%$ are missing) than for the neutron (about $44\,\%$). There is no simple explanation of such a large discrepancy, but it certainly calls for further measurements of $\Delta\sigma$ and corresponding phenomenological studies in the region of $E_\gamma$ beyond the one covered by existing data --- in particular because of the apparently divergent behavior of $\sigma_\mathrm{tot}$ 
beyond $\sqrt{s}$ of a few times $10~\mathrm{GeV}$.  
In this energy domain, processes can be studied in a Regge approach, 
in which the $s$-dependence of $\Delta\sigma$ for either real or virtual 
polarized photoabsorption is of the form~\cite{Bass:1997fh}
\begin{equation}
    \Delta\sigma 
    = \left[ I c_1 s^{\alpha_{a_1}-1} + c_2 s^{\alpha_{f_1}-1} 
    + c_3 \frac{\log s}{s} + \frac{c_4}{\log^2s} \right] F(s,Q^2) \>,
    \label{eq:DS}
\end{equation}
where $I = \pm 1$ is the isospin factor corresponding to the proton or neutron,
respectively, $\alpha_{a_1}$ and $\alpha_{f_1}$ are the Regge intercepts of the $a_1$ and $f_1$ Regge trajectories, and $Q^2$ is the negative square of the virtual photon four-momentum. For photoproduction (real photons) $Q^2 = 0$, the logarithmic terms are negligible and $F(s,Q^2)$ simplifies to a constant that can be absorbed in the $c_1$ and $c_2$ coefficients. The Regge parameterization of $\Delta\sigma$ then becomes
\begin{equation}
    \Delta\sigma = I {c}_1 s^{\alpha_{a_1}-1} + {c}_2 s^{\alpha_{f_1}-1} \>.
    \label{eq:D6}
\end{equation}
A commonly accepted set of parameters is ${c}_1 = (-34.1\pm 5.7)\,\mu\mathrm{b}$, $\alpha_{a_1} = 0.42\pm 0.23$, ${c}_2 = (209.4\pm 29.0)\,\mu\mathrm{b}$ and  $\alpha_{f_1} = (-0.66\pm 0.22)$~\cite{Helbing:2006zp}.  

Since the proton and neutron anomalous magnetic moments,
$\kappa_p$ and $\kappa_n$, can be decomposed into their
isovector and isoscalar components, $\kappa_p = (\kappa_\mathrm{s} + \kappa_\mathrm{v})/2$
and $\kappa_n = (\kappa_\mathrm{s} - \kappa_\mathrm{v})/2$, hence
\begin{equation}
    \kappa_{p,n}^2 
    = \textstyle{\frac{1}{4}}\kappa_\mathrm{s}^2
    \pm \textstyle{\frac{1}{2}}\kappa_\mathrm{v}\kappa_\mathrm{s}
    + \textstyle{\frac{1}{4}}\kappa_\mathrm{v}^2 \>,
    \label{eq:D8}
\end{equation}
the GDH sum rule can accordingly be split into its ``scalar-scalar'', ``vector-vector'' 
and ``vector-scalar'' parts ($I_\mathrm{GDH}^\mathrm{ss}$, $I_\mathrm{GDH}^\mathrm{vv}$, 
and $I_\mathrm{GDH}^\mathrm{vs}$, respectively)~\cite{Karliner:1973em}. In particular,
\begin{equation}
I_\mathrm{GDH}^\mathrm{vs} 
    = \int_{E_\gamma^\mathrm{thr}}^\infty
    \left( \sigma_{3/2}^\mathrm{vs} - \sigma_{1/2}^\mathrm{vs} \right) \frac{dE_\gamma}{E_\gamma}
    = {\textstyle{\frac{1}{2}}} \kappa_\mathrm{v}\kappa_\mathrm{s} \frac{2\pi^2\alpha}{M^2} \>.
\end{equation}
Since $\kappa_p^2 - \kappa_n^2 = \kappa_\mathrm{v}\kappa_\mathrm{s}$, 
the isovector GDH sum rule involving the {\sl difference\/} of 
$\Delta\sigma$ for proton and neutron targets, 
$\Delta\sigma_{p-n} \equiv \Delta\sigma_p - \Delta\sigma_n$, 
predicts
\begin{equation}
    \int_{E_\gamma^\mathrm{thr}}^\infty
    \frac{\Delta\sigma_{p-n}}{E_\gamma} \, dE_\gamma
    = 2 I_\mathrm{GDH}^\mathrm{vs} 
    \approx -27.5\,\mu\mathrm{b} \>.
    \label{eq:deltasigpn}
\end{equation}
Eq.~(\ref{eq:deltasigpn}) should also provide the $Q^2$-derivative of the famous generalized Bjorken sum rule~\cite{Bjorken:1966jh} in the $Q^2 \to 0$ limit~\cite{Deur:2018roz}. This is borne out by the low-$Q^2$ measurements of the Bjorken sum rule performed at JLab~\cite{Deur:2004ti}.
The $Q^2=0$ result for the latter is obtained from a fit of the data that is 
extrapolated to $Q^2=0$ without constraint from chiral effective field theory.

\begin{figure}[ht]
\vspace{0.4cm}
\centering
{
    \includegraphics[width=0.4\textwidth,angle=90,keepaspectratio]{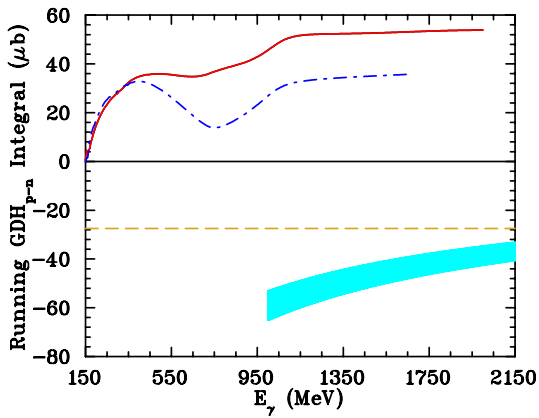}
}
\caption{Running isovector GDH integral.  The red solid and blue dash-dotted lines represent 
the SAID MA19 \protect\cite{A2:2019yud} and MAID2007 \protect\cite{Drechsel:2007if} solution, respectively.  The yellow dashed line indicates the predicted limiting value as given by Eq.~(\ref{eq:deltasigpn}). 
The cyan band shows the Regge estimate of the complement to the running integral from the current energy to infinity.}
\label{fig:GDHiso}
\end{figure}

In Regge theory, $\Delta\sigma_{p-n}$ is driven by the $a_1$ trajectory alone,
\begin{equation}
    \Delta\sigma_{p-n}^\mathrm{Regge} 
    = 2 {c}_1 s^{\alpha_{a_1}-1} \>,
    \label{eq:D7}
\end{equation}
and one can visualize its contribution by evaluating the running integral 
of (\ref{eq:D7}) divided by $E_\gamma$ from a chosen {\sl lower\/} integration limit to infinity.  
The Regge contributions to the running isovector GDH integrals are shown by cyan bands in Fig.~\ref{fig:GDHiso}.  As in Fig.~\ref{fig:GDHrun}, these (negative) running values need to be added to the (positive) SAID and MAID curves, apparently bringing the totals closer to the predicted value.

The energy dependence of $\Delta\sigma_{p-n}$ in Regge theory is also interesting by itself.  If one makes the usual assumption that the $a_1$ trajectories are straight lines parallel to the $\rho$ and $\omega$ trajectories, one finds $\alpha_{a1} \approx -0.4$ for the leading trajectory, with the sign opposite to the one obtained by analyzing deeply inelastic scattering, photo- and electroproduction data.  Understanding the sign and the magnitude of the Regge intercept $\alpha_{a1}$ is important, as this parameter is intimately connected to the quark string tension of QCD: if $\alpha'$ is the corresponding reggeon slope, the intercept 
is given by $\alpha_{a1} = 1 - \alpha' m_{a1}^2$, and $\alpha' = 1/(2\pi\sigma_q)$, where $\sigma_q$ is the string tension \cite{Donnachie02}. A new determination of these parameters by extending 
the $\Delta\sigma$ measurements both on proton and neutron (deuteron) targets to $E_\gamma \approx 12~\mathrm{GeV}$ is one of the aims of the recently approved REGGE Experiment in Hall~D of Jefferson Lab~\cite{Dalton:2020wdv}.


\section{Conclusion}

The evaluation of three sum rules (GDH, Baldin, and GGT) involving the difference of helicity-dependent total photoabsorption cross-sections, $\Delta\sigma$, shows that single-pion photoproduction off the nucleon is the dominant contribution to these sum rules.  In all cases, the single-pion contribution converges above $E_\gamma \approx 1.7~\mathrm{GeV}$.  The situation is the most favorable for the Baldin sum rule where --- to the attainable levels of precision --- the single-pion photoproduction off the nucleon comes closest to agreeing with Compton scattering and Effective Field Theory calculations, in particular in the neutron case.  In the GDH sum rule for the proton, the single-pion contribution saturates the sum rule to about 90\%, while in the neutron case, the missing strength amounts to about 44\%.  In the case of the GGT sum rule, the lack of precise calculations and the size of experimental uncertainties preclude a clear statement of agreement.

\vspace{0.5cm}
\section*{Acknowledgments}

We thank Lothar Tiator, Mikhail Bashkanov, Michael Eides, Harald Grie{\ss}hammer, Viktor Kashevarov, Vladimir Pascalutsa, and Paolo Pedroni for useful discussions.
This work was supported in part by the U.~S.~Department of Energy, Office of Science, Office of Nuclear Physics, under Awards No.~DE--SC0016583 and No.~DE--SC0016582, and in part by the U.S. Department of Energy, Office of Science, Office of Nuclear Physics under contract DE-AC05-06OR23177.


\end{document}